\documentclass[sigconf]{acmart}
\AtBeginDocument{%
  }

\usepackage{booktabs}
\usepackage{graphicx}
\usepackage{subcaption}
\usepackage{cleveref}
\usepackage{hyperref}
\usepackage{xcolor}
\usepackage{tikz}

\usepackage{graphicx}
\newcommand{\orangeicon}{\raisebox{-0.2\height}{\includegraphics[height=1.2em]{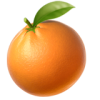}}}

\usepackage{xspace}
\usepackage{stfloats}
\usepackage[most]{tcolorbox}
\usepackage{framed}
\usepackage{xcolor}
\usepackage{fontawesome5}

\newcommand{\gptatlas}{\includegraphics[height=1em]{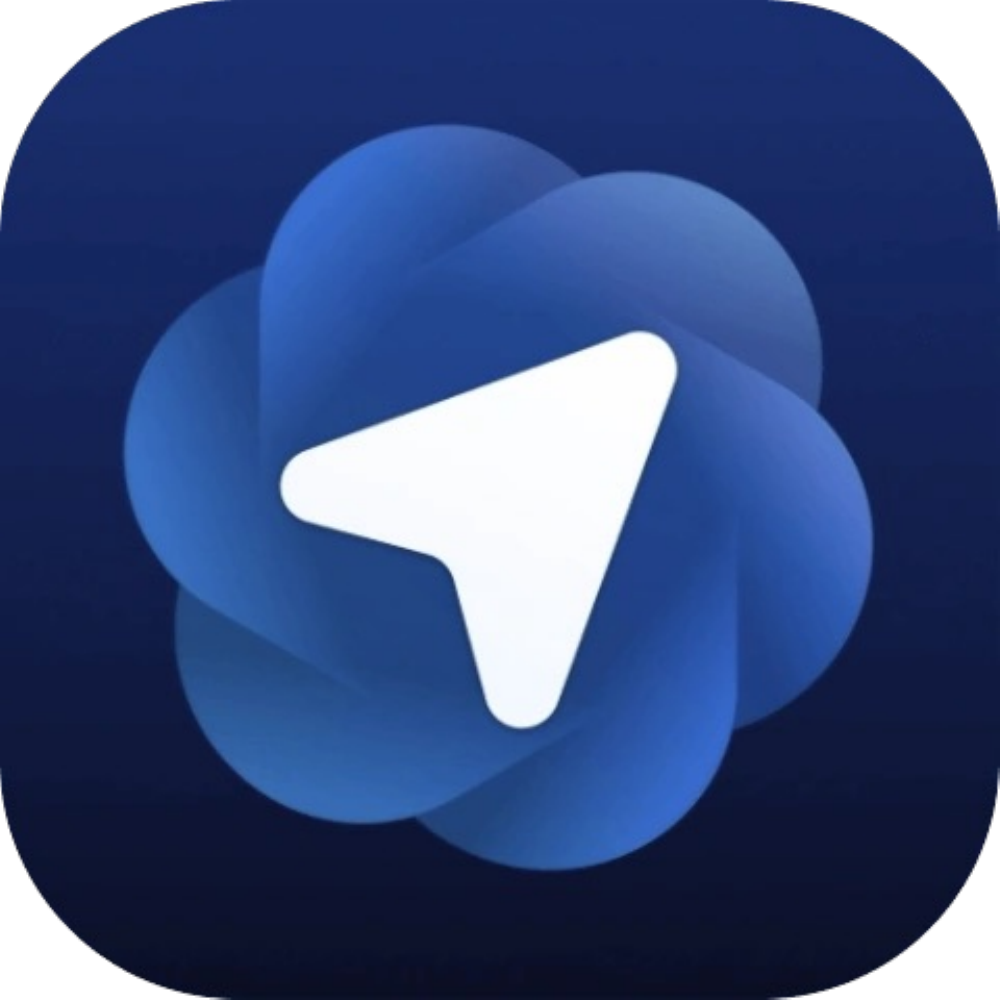}}

\setcopyright{none}
\renewcommand\footnotetextcopyrightpermission[1]{}
\acmConference{}{}{}

\definecolor{ForestGreen}{RGB}{34, 139, 34}

\newcommand{\increasenoparent}[1]{\textcolor{ForestGreen}{+#1}}

\newcommand{\subsec}[1]{\noindent\textbf{#1}~~}
\newcommand{\webhighlight}[2][yellow!55]{\colorbox{#1}{\small\ttfamily #2}}
\newcommand{\hotcite}[3][yellow!55]{\webhighlight[#1]{[#2:\textquotedbl{}#3\textquotedbl{}]}}

\newcommand{\hidebox}[1]{\fcolorbox{red!75!black}{white}{\small\ttfamily #1}}

\definecolor{xwebguidegreen}{HTML}{2ed573}
\definecolor{xwebguidered}{HTML}{ff4757}

\definecolor{xweblink}{HTML}{5B6EE1}
\definecolor{xwebcyan}{HTML}{31529F} % soft teal
\definecolor{xwebbackground}{HTML}{CFDFF6}
\definecolor{xwebpink}{HTML}{e84393}
\definecolor{customviolet}{HTML}{7950F2}

\definecolor{findarrowcolor}{HTML}{FFAB40}
\definecolor{guidearrowcolor}{HTML}{0D5BDC}
\definecolor{hidearrowcolor}{HTML}{7950F2}

\newcommand{\Apref}[1]{Appendix~\ref{#1}}

\newcommand{\agent}{\textsc{PageGuide}\xspace}
\newcommand{\FIND}{\textsc{Find}\xspace}

\newcommand{\GUIDE}{\textsc{Guide}\xspace}

\newcommand{\eg}{\textit{e.g.,}\xspace}
\newcommand{\ie}{\textit{i.e.,}\xspace}

\newcommand{\modeicon}{\hspace{0em}\raisebox{-1pt}{\includegraphics[scale=0.04]{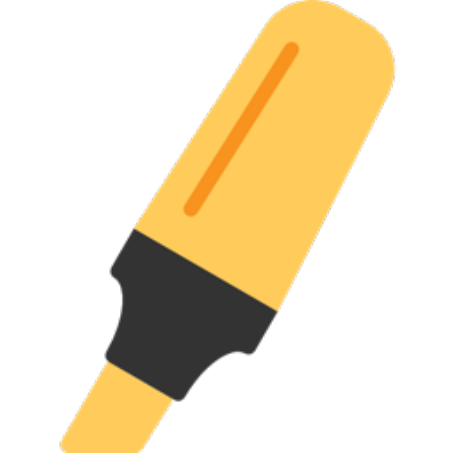}}\xspace}
\newcommand{\evidenceicon}{\hspace{0em}\raisebox{-1pt}{\includegraphics[scale=0.04]{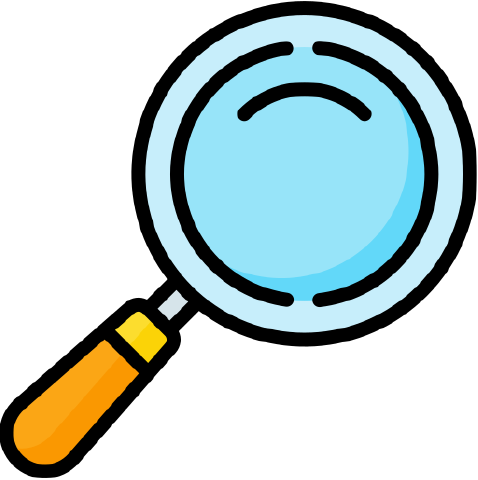}}\xspace}
\newcommand{\taskicon}{\hspace{0em}\raisebox{-1pt}{\includegraphics[scale=0.04]{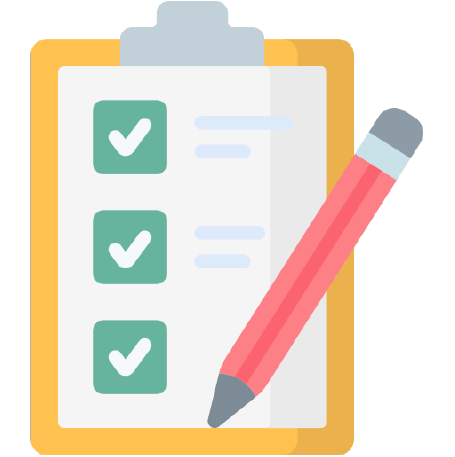}}\xspace}
\newcommand{\labelicon}{\hspace{0em}\raisebox{-1pt}{\includegraphics[scale=0.04]{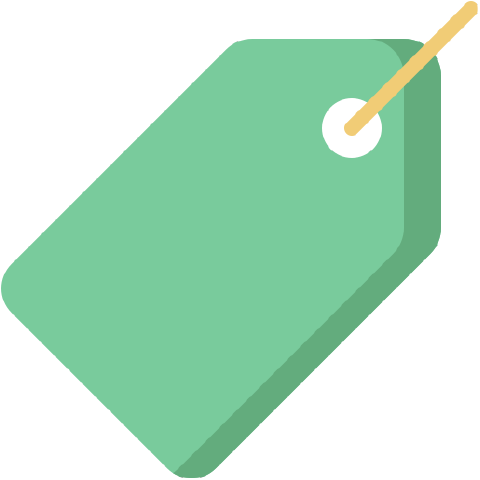}}\xspace}

\newcommand{\users}{\textsc{92}\xspace}
\newcommand{\projectrepo}{\href{https://pageguide.github.io}{PageGuide.github.io}\xspace}

\begin{document}

%% The "title" command has an optional parameter,
%% allowing the author to define a "short title" to be used in page headers.
\title{\texorpdfstring{\orangeicon\ }{} \agent: Browser extension to assist users in navigating a webpage and locating information}

%%
%% By default, the full list of authors will be used in the page
%% headers. Often, this list is too long, and will overlap
%% other information printed in the page headers. This command allows
%% the author to define a more concise list
%% of authors' names for this purpose.
\renewcommand{\shortauthors}{Nguyen et al.}

%%
%% Authors
\author{Tin Nguyen}
\authornote{Major contributions. Thang T. Truong collected the dataset and built the automated testing pipeline. Runtao Zhou conducted the user study.}
\authornote{Tin Nguyen is the team lead who led the core implementation, user study design, and team coordination.}
\affiliation{%
  \institution{Auburn University}
  \city{Auburn}
  \state{Alabama}
  \country{USA}
}
\email{ttn0011@auburn.edu}

\author{Thang T. Truong}
\authornotemark[1]
\affiliation{%
  \institution{Auburn University}
  \city{Auburn}
  \state{Alabama}
  \country{USA}
}
\email{tht0021@auburn.edu}

\author{Runtao Zhou}
\authornotemark[1]
\affiliation{%
  \institution{University of Virginia}
  \city{Charlottesville}
  \state{Virginia}
  \country{USA}
}
\email{uar6nw@virginia.edu}

\author{Trung Bui}
\affiliation{\institution{}
  \city{}
  \state{}
  \country{}
}
\email{bhtrung@gmail.com}

\author{Chirag Agarwal}
\affiliation{%
  \institution{University of Virginia}
  \city{Charlottesville}
  \state{Virginia}
  \country{USA}
}
\email{chiragagarwal@virginia.edu}

\author{Anh Totti Nguyen}
\affiliation{%
  \institution{Auburn University}
  \city{Auburn}
  \state{Alabama}
  \country{USA}
}
\email{anh.ng8@gmail.com}

\begin{abstract}
  Users browsing the web struggle to locate relevant information on cluttered pages and to complete web navigation tasks. Current web agents can answer questions and automate actions, but return answers without showing where the information comes from, forcing users to manually verify results and blindly trust every automated step. We present \textbf{\agent}, a browser extension that grounds LLM answers in page content (\eg HTML DOM elements, text, maps, images, charts), via two features: \textbf{\FIND}, which highlights supporting evidence in situ on a page, and \textbf{\GUIDE}, which guides users across pages with evidence captured at each step. In a within-subject study ($N{=}\users$), \agent improves overall verification accuracy from 65\% to 77\% and reduces verification time from 73s to 67s across \FIND and \GUIDE tasks. \agent also reduces manual search effort and improves users' perceived support for verification. Code is available at \projectrepo.

\end{abstract}

%%
%% The code below is generated by the tool at http://dl.acm.org/ccs.cfm.
%% Please copy and paste the code instead of the example below.


%%
%% Keywords. The author(s) should pick words that accurately describe
%% the work being presented. Separate the keywords with commas.
%% A "teaser" image appears between the author and affiliation
%% information and the body of the document, and typically spans the
%% page.
\begin{teaserfigure}
    \centering
    \includegraphics[width=0.95\textwidth]{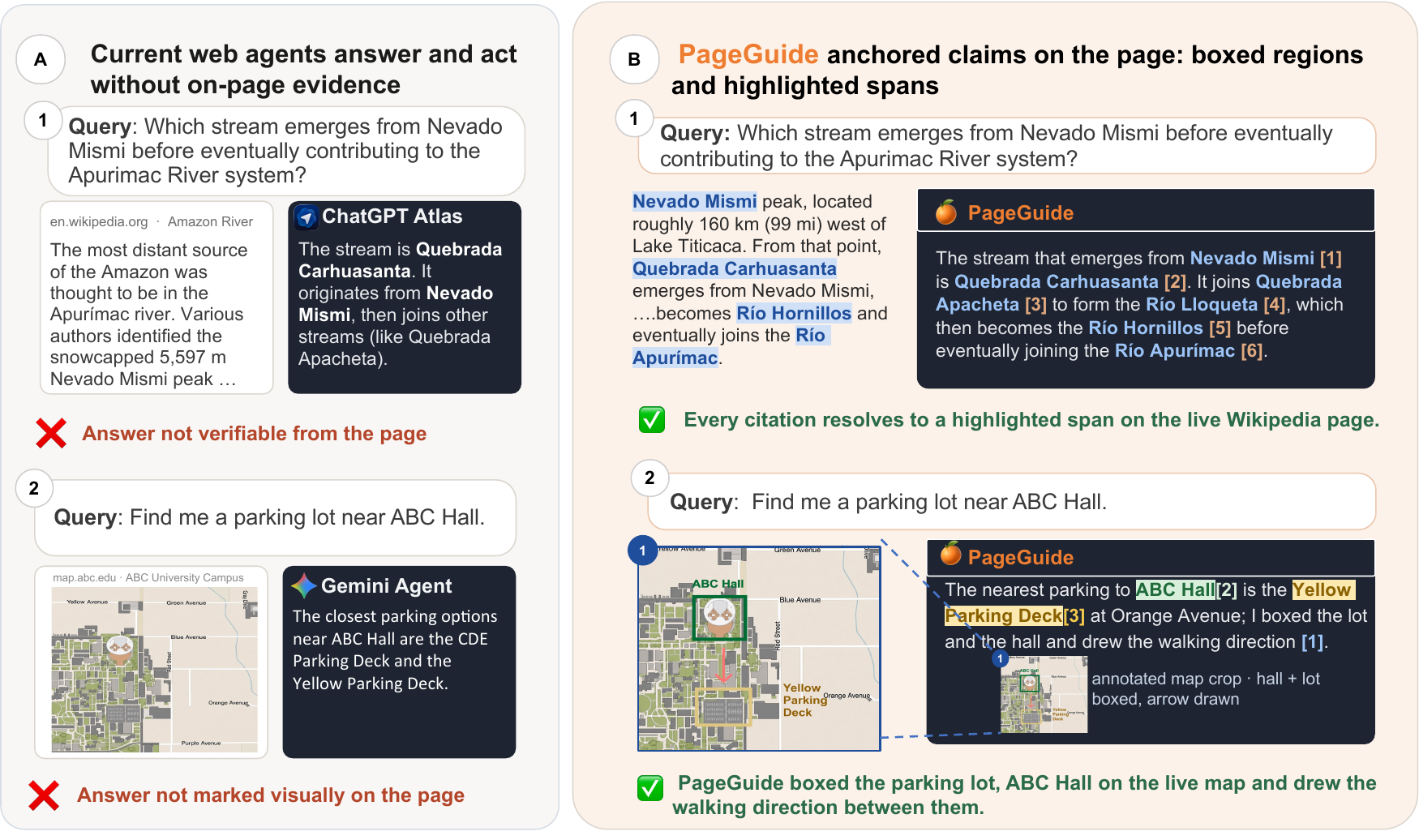}
    \caption{Existing web agents provide limited support for user verification. For example, on the \href{https://en.wikipedia.org/wiki/Amazon_River}{Wikipedia page} (A1), ChatGPT Atlas {\protect\gptatlas} returns the correct answer, but it does not highlight the supporting evidence on the page, making the response difficult for the user to verify. In contrast, in (B1), \agent shows the answer with clickable references and highlights the corresponding evidence on the page (\eg clicking Nevado Mismi\textsuperscript{[1]} jumps to the highlighted Nevado Mismi on the page). Similarly, in (A2), on \href{https://www.map.abc.edu}{Map}, Gemini Agent (\protect Gemini 3.0 Flash) cannot point to the parking lot's location, making it difficult for users to locate it on the dense map. In contrast, in (B2), \agent provides visual grounding by annotating both the response and the page: it boxes the target hall and parking lot and draws the relation between them, allowing users to directly locate and verify the answer on the map.}
    \label{fig:three_figures_find_example}
    \Description{Comparison of web-agent answers with and without on-page evidence. Two columns pair text and map examples. On the left, answers appear separately from the source content. On the right, numbered PageGuide citations correspond to highlighted text spans in the Wikipedia example and boxed locations in the map example. Dashed connectors link a map region to an enlarged annotated crop in the answer.}
\end{teaserfigure}

% \received{20 February 2007}
% \received[revised]{12 March 2009}
% \received[accepted]{5 June 2009}

%%
%% This command processes the author and affiliation and title
%% information and builds the first part of the formatted document.
\maketitle

\section{Introduction}
\label{sec:introduction}
AI chatbots have become a primary tool for information retrieval, with 62.8\% of the internet users citing information retrieval as a top reason for going online \citep{datareportal2025global}. In WebArena, VisualWebArena, and WebVoyager \citep{webarena,visualwebarena,webvoyager}) platforms, chatbots read the page content, autonomously execute actions (\eg clicking buttons and links or filling out forms), and return answers in a chat sidebar. However, modern webpages often contain important information beyond text represented in the HTML DOM, including maps, images, charts, canvases, and other visually rendered elements. Accordingly, in VisualWebArena, many tasks require agents to reason not only over textual page content, but also over visually rendered information such as images, maps, and interface layouts. On these tasks, web agents achieve a success rate of only 14.9\%, highlighting the difficulty of visual reasoning and the need for users to verify the evidence underlying an agent's answers and actions. However, current agents provide limited support for such verification: they return \textbf{answers without referencing where they are extracted from, and act without pausing for user verification}. 
Users must blindly trust every automated step and manually cross-check answers against a webpage---making it challenging to inspect, correct, or intervene (see \Cref{fig:three_figures_find_example}a).

% Existing web agents do not address this problem. 
Commercial LLM assistants (\eg RealWebAssist \citep{yerealwebassist}, Browser-Use \citep{browseruse2025}, ChatGPT Atlas \citep{atlas2025}, Dia Browser \citep{diabrowser2025}) and web agents (\eg Claude in Chrome \citep{anthropic_claude_chrome_2026}, Gemini in Chrome \citep{geminichrome2025}, OpenAI Operator \citep{operator2025}) optimize for end-to-end web navigation but make no attempt to ground their outputs in the page or expose intermediate decisions to the user. Grounded reasoning approaches such as HoT \citep{highlightedcot} and ScreenAI \citep{baechler2024screenai} link model outputs to evidence spans, but do so within the model’s response text---the webpage itself is left unchanged, so users still have to locate the evidence themselves. Standard keyword search (\eg Ctrl+F) is limited to exact string matching and cannot answer questions or highlight supporting evidence in context. None of these approaches give users a way to verify answers or confirm actions on the page.

Two specific user needs remain unaddressed. \textbf{(1) Finding and verifying information} on dense, cluttered pages is tedious: users must scan long pages manually to find where AI answers may be generated based on \citep{10.1145/3742413.3789134}. \textbf{(2) Completing navigation tasks with step-level oversight} is risky without step-level oversight \citep{kuntz2025harm,zhang2024agent,levy2024st,tur2025safearena,jones2026benign,cuvin2025decepticon,rwom2025,cowpilot2025}: users often need to follow many steps (e.g., clicking, scrolling, navigating menus) to complete tasks such as \textit{Look up The Lotte Hotel Seoul and The Joseon Hotel Seoul on Booking.com and report rating + price/night. Then on Google Flights, find the cheapest London→Seoul round-trip in mid-July–early August. Pick the best overall plan}. However, agent's instructions are long and do not point to where the relevant information are located on the page.

\begin{figure*}[ht]
    \centering
    \includegraphics[width=0.95\textwidth]{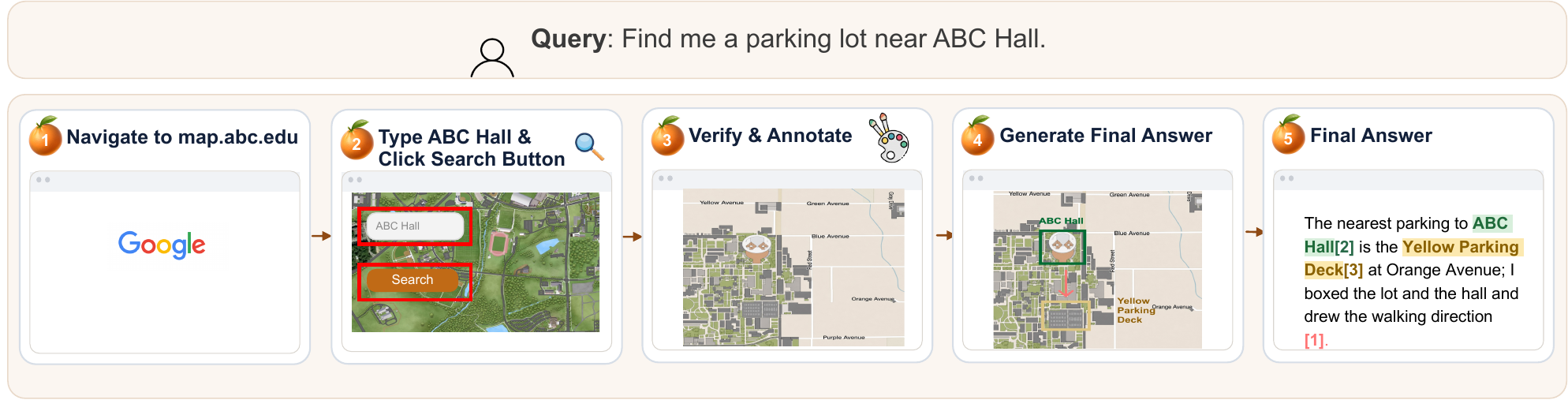}
    \caption{Example of a \GUIDE process of \textit{finding a parking lot} on a map. Given the query, \agent generates a step-by-step plan and presents one instruction at a time. At each step, the target HTML DOM element is highlighted directly on the page (\eg \hidebox{Search}), while the sidebar displays the corresponding instruction. To capture the evidence on the map canvas, \agent has to capture the image region and annotate using boxes, and arrows to make the evidence visible to users. Finally, \agent provides the final answer with clickable references; selecting a reference navigates the page directly to the corresponding DOM evidence or reveals the annotated image region.}
    \label{fig:guide_example}
    \Description{A five-stage diagram shows PageGuide locating parking near a hall. Arrows connect the stages from left to right: open the map website, enter the hall name and select Search, inspect the map, annotate the hall and parking location, and present a final answer. Boxes identify the search controls and map locations; numbered references connect the final answer to the evidence.}
\end{figure*}

We present \textbf{\agent}, a browser extension that addresses these needs by grounding LLM answers directly in both HTML DOM elements and non-HTML DOM visual content (\eg maps, images, canvases, and charts), using visual overlays such as bounding boxes, arrows, and stars.
\agent offers two main features: (1) \textbf{\FIND} links model answers to specific DOM or non-DOM elements in the page via inline citations, highlighting supporting evidence in-place so users can instantly verify any claim without leaving the page (\Cref{fig:three_figures_find_example}b); (2) \textbf{\GUIDE} guides users through a task step by step, capturing and highlighting the relevant evidence for each intermediate action or sub-task. Users can pause the agent at any step to inspect the evidence and verify the action before proceeding, following a mixed-initiative copilot model \citep{cowpilot2025,amershi2019guidelines,horvitz1999mixedinitiative,sweller1988cognitive} (\Cref{fig:guide_example}).

To evaluate \agent, we conduct a within-subject controlled user study ($N{=}\users$) on websites (\Cref{sec:data_collection}), where each participant completes eight tasks spanning all combinations of grounding condition (grounded vs. non-grounded), interaction mode (\FIND vs. \GUIDE), evidence type (text vs. visual), and answer correctness (correct vs. incorrect) (see \Cref{fig:user_study_display}). We measure task completion time, accuracy, mouse hovering signal, and post-task Likert ratings. Our study aims to answer three central research questions:

\noindent\textbf{\textit{RQ1:}} Does \agent improve verification accuracy across all \FIND and \GUIDE?

\noindent\textbf{\textit{RQ2:}} How does grounding affect verification time and interaction effort across \FIND and \GUIDE tasks?

\looseness=-1\noindent\textbf{\textit{RQ3:}} Do users perceive completing tasks with \agent as easier and less effortful compared to completing the same tasks without grounding?

By systematically comparing two conditions (non-grounded and grounded), we aim to provide rigorous empirical evidence on whether grounding LLM outputs leads to measurable improvements in task efficiency, accuracy, and user experience across both interaction features.

\noindent In summary, this paper makes \textbf{three contributions}:
\begin{enumerate}
    \item A unified \textit{\FIND--\GUIDE} framework that supports two complementary interaction modes: \FIND locates information directly on a specific page, while \GUIDE supports information seeking across pages through navigation.
    \item A controlled user study ($N{=\users}$ participants) shows that grounded, verifiable web interaction improves both verification accuracy and time across \FIND and \GUIDE tasks.
    \item We discuss how grounded answer design affects perceived ease of use, confidence, and sense of control in AI-assisted browsing, drawing on post-study questionnaire results across \FIND and \GUIDE.
\end{enumerate}

\section{Related Work}
\label{sec:related_work}
\looseness=-1 Web agents are built to serve a purpose: \textit{helping users accomplish goals on the web more efficiently and accurately than they could alone}. Yet the dominant design paradigm (\ie fully autonomous agents that plan, click, and return a final result) conflates \textit{completing a task} with \textit{serving the user}. When a user needs to verify a claim on a cluttered page, follow a multi-step task, or stay focused amid irrelevant content, end-to-end automation offers little help. Prior work can be understood through two orthogonal design choices that determine how well a system actually serves users in these situations: \textit{(i)~autonomy level}, whether the system acts on behalf of the user or alongside them; and \textit{(ii)~output grounding}, whether the answer is returned without showing where it comes from on the page or linked to specific, visible elements. Systems that score high on autonomy and low on grounding (most existing agents) maximize speed at the cost of verifiability. \agent inverts this trade-off: it acts as a copilot that grounds every output (\eg answers, or actions) directly in the HTML DOM, text spans, canvas, tables, charts, or images keeping the user in control at every stage.\\

\noindent\textbf{Web Agents and Browser Automation.} 
LLM-based web agents have made rapid progress on benchmarks that measure end-to-end task success \citep{webarena,visualwebarena,webvoyager,mind2web}. Early systems such as Auto-GPT \citep{autogpt2023} and BabyAGI \citep{babyagi2023} decompose high-level goals into sub-tasks, execute them via code or tool calls, and iterate until done. More recently, browser-integrated systems---OpenAI Operator \citep{operator2025}, Claude in Chrome \citep{claudecomputeruse2024}, Gemini in Chrome \citep{geminichrome2025}, Browser Use \citep{browseruse2025}, Dia Browser \citep{diabrowser2025}, and ChatGPT Atlas \citep{atlas2025}---embed LLM planners directly in the browser, enabling click, type, scroll, and navigate actions on pages. Synapse \citep{jin2022synapse}, iTutor \citep{zou2023itutor}, and HelpViz \citep{zhong2021helpviz} generate interactive tutorials, but are limited to a fixed set of demonstrations and supported tasks. Cowpilot \citep{cowpilot2025} proposes actions and the user is able to pause, reject, or take alternative action when needed. Morae \citep{peng2025morae} also has the agent proactively decide \textit{when} to pause for clarifying users's intent; however, their confirmation step do not ground the proposed action in visible page evidence, making each approval harder for the user to verify. Complementarily, \textit{When Should Users Check?} \citep{zhou2026should} studies the \textit{frequency and timing of confirmation} in multi-step agent tasks. Rather than confirming only at the end or interrupting after every action, it models where intermediate confirmations should be placed to balance the cost of user interruptions against the cost of detecting an error late and repeating subsequent steps. These works address when an agent should involve the user and
largely assume that users remain engaged during task execution so they can respond to interruptions or confirmations as they arise. In real-world use, however, users may either continuously monitor the agent or delegate a task, shift their attention elsewhere, and return
after the agent has made progress or completed the task. All of these systems sit at the \textit{high-autonomy, low-grounding} end of the design space: they handle the task for the user and return a final result, but their intermediate decisions are opaque and they do not show where the output comes from on the page.

This opacity carries real costs. Recent benchmarks show that autonomous web agents often make incorrect decisions in high-stakes situations, such as deleting data, making unintended purchases, or following deceptive “Download” buttons that expose personal information \citep{kuntz2025harm,zhang2024agent,levy2024st,tur2025safearena,cuvin2025decepticon}. Even in everyday tasks, stale or visually embedded information can lead to incorrect answers. For example, a recreation center may frequently update its opening and closing hours in a visual timetable, yet an agent may rely on outdated information and report an incorrect closing time. Grounding the response in the currently displayed timetable would allow users to quickly verify that the reported hours reflect the latest information. A partial remedy is the copilot model named Cowpilot \citep{cowpilot2025}, where the agent proposes actions and the user is able to pause, reject, or take alternative action when needed. \agent extends this idea by not only preserving user control but also making it easier for users to verify the agent's actions and outputs.\\

\noindent\textbf{Chain-of-Thought and Grounded Reasoning.} Chain-of-Thought (CoT) prompting improves LLM accuracy by externalizing intermediate reasoning steps \citep{wei2022cot,agarwal2024faithfulness,tanneru2024difficulty,lobo2025impact}. Highlighted Chain-of-Thought (HoT) \citep{highlightedcot} takes this further by embedding XML-style evidence tags in each reasoning step, linking conclusions to specific input spans so users can trace \textit{why} an answer is produced. However, HoT must regenerate the cited input alongside its answer, which works well for short-context inputs but cannot scale to long, cluttered webpages, since regenerating the full page within the response is impractical. Vision-language work on web interfaces reinforces this direction: ScreenAI \citep{baechler2024screenai} and WebQuest \citep{wang2024webquest} show that reliable web interaction requires aligning model outputs with screen elements and multimodal page evidence, not just fluent text.

These approaches do not help users verify answers or actions directly on the webpage. \agent closes this gap by moving evidence directly onto the page: instead of returning ``the answer is supported by element 42,'' it places a highlight \textit{on} element 42 in the HTML DOM, making the evidence visible exactly where the user is already looking.\\

Taken together, \agent enables users to verify its answers and intermediate steps by directly surfacing textual and visual evidence, particularly in long-horizon, multi-step tasks.

\section{Method}
\label{sec:method}

\subsection{System Overview}
\label{sec:system_overview}
\begin{figure*}
    \centering
    \includegraphics[width=0.95\textwidth]{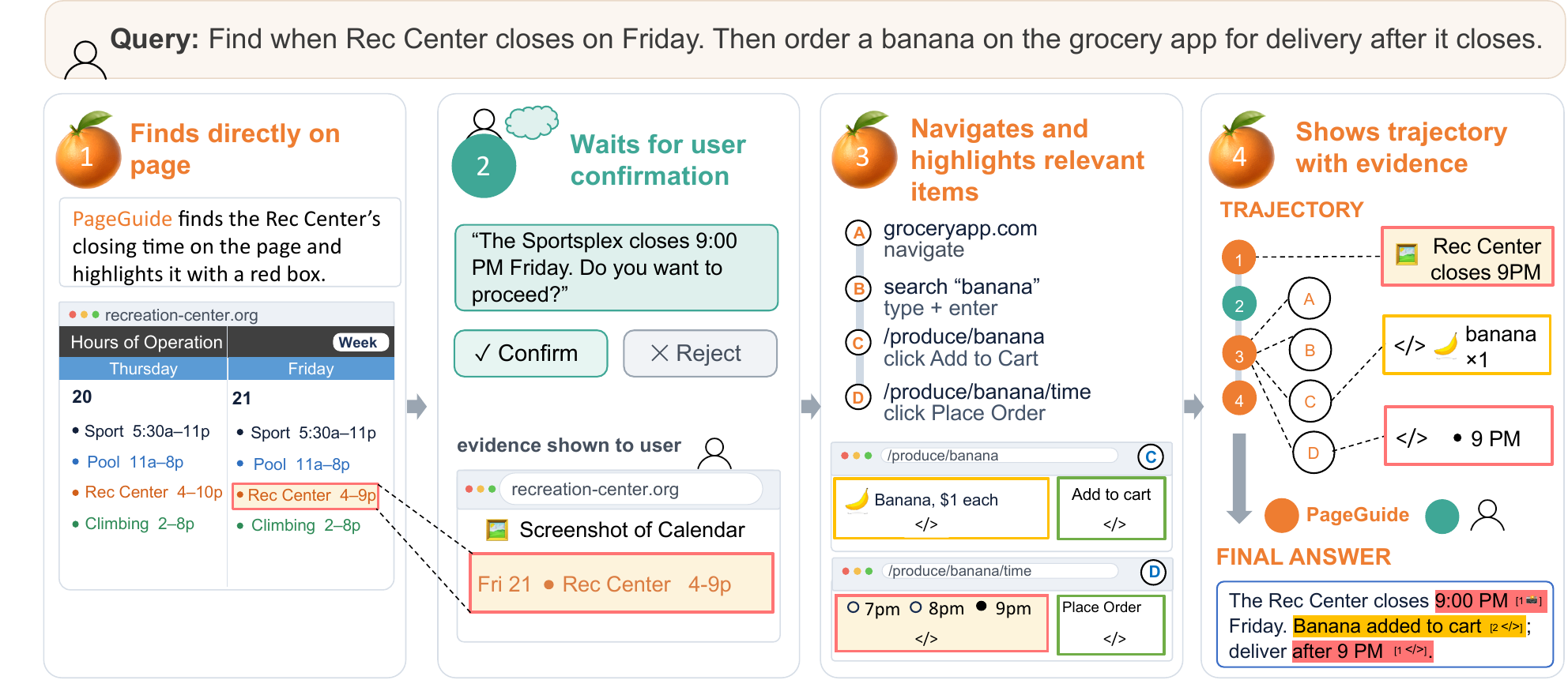}
    \caption{\agent executes the task while keeping users in the process and showing verifiable evidence throughout the trajectory. Given the query \textit{``Find when Rec Center closes on Friday. Then order a banana on the grocery app for delivery after it closes.''}, (1) \agent first finds the Rec Center's Friday closing time and highlights the supporting evidence on the page. (2) Before proceeding, \agent pauses for user confirmation. (3) After approval, \agent navigates the grocery app, selects a banana, and schedules delivery for a valid time after the Rec Center closes. (4) \agent presents the complete trajectory together with the textual and visual evidence supporting each step, enabling the user to verify the final answer.}

    \label{fig:overview}
    \Description{A four-stage workflow connects evidence lookup, user confirmation, navigation, and an evidence-linked task history. Arrows lead from a highlighted recreation-center closing time to Confirm and Reject controls, then to grocery search, cart, and delivery controls. The final panel connects numbered actions to supporting text or screenshots and includes a cited answer. Dashed lines show which evidence supports each action.}

\end{figure*}
\agent is a Manifest V3 Chrome extension that injects content scripts into webpages, giving it both read and write access to the HTML DOM in real time. Our design is guided by three goals: \textbf{(1)} surface supporting evidence both during agent execution and after task completion, \textbf{(2)} ground the agent's answer claims in corresponding page evidence (\eg text spans, images, charts) to support verification, and \textbf{(3)} flexibly adapt the interaction to the user's intent through \FIND and \GUIDE modes. As shown in Figure~\ref{fig:overview}, the system follows a pipeline: the \textbf{Router} classifies the user's query and dispatches it to the appropriate handler (\FIND or \GUIDE); each handler \textbf{reads the page content} to ground its answers and evidence in the page, then highlights relevant evidence and surfaces step-by-step instructions. The two handlers differ in how they interact with users:

\begin{itemize}
    \item \textbf{\FIND} is single-pass: the agent identifies supporting evidence spans and renders inline highlights on the page in one shot.
    \item \textbf{\GUIDE} is iterative: the agent proposes one action at a time (see \Cref{fig:overview}) and will capture the evidence during the navigation process.
\end{itemize}

\subsection{Reading the Page State}
Before the Router or any handler can reason about the page, \agent must convert the HTML DOM into a form the LLM can work with. Using a Set-of-Marks (SoM) \citep{visualwebarena} style approach, every visible, or interactive node is assigned a unique integer index, producing a structured element index $\mathcal{D} = \{(id_j,\, \text{text}_j,\, \text{tag}_j,\, \text{bbox}_j)\}_{j=1}^{m}$ that records each element's content, tag type, and on-screen bounding box. This index is passed to the Router for intent classification and to whichever handler is invoked, enabling the LLM to refer to specific page elements by index rather than parsing raw HTML. For \GUIDE, the HTML DOM is re-read after each user-confirmed action so that the next step reflects the current page state.

\subsection{Intent Router}
\FIND and \GUIDE support different user intents and therefore require different interaction strategies. \FIND is designed for queries that ask the agent to locate and ground relevant evidence on the current page, whereas \GUIDE handles multi-part queries that require navigation across several steps while capturing evidence for each subtask. Because of that, \agent first uses an Intent Router to determine which mode is appropriate before any page-level interaction begins.

Once the HTML DOM index $\mathcal{D}$ is ready, the Router is the first component the query reaches \Cref{fig:overview}. Its sole job is to decide which of the two handlers should run before any page-level interaction begins. Given query $q$ and a compact page context $\mathcal{C}$ (\eg page title, content type), the router selects:
\begin{equation}
    \text{mode} = f_{\text{router}}(q, C), \text{where~~~mode} \in \{\text{\FIND, \GUIDE}\}.
\end{equation}

The router is a single LLM call with a structured prompt that describes each mode with canonical examples and asks for a mode label plus a one-sentence justification (see full prompt in \Cref{fig:routing_prompt} in Appendix). Routing by mode allows each pathway to be optimized for its specific task, rather than relying on a single monolithic prompt to handle all cases.

\noindent\textbf{Router Prompt (condensed)}
\begin{framed}
\textbf{System:} Given a user query and a brief page context, classify the query into one of two features: FIND (factual lookup), and GUIDE (step-by-step task). Return the mode label and a one-sentence justification.\\[2pt]
\textbf{User:} Query: \textit{``\{query\}''} \quad Page context: \textit{\{page\_title, content\_type\}}
\end{framed}
% \begin{tcolorbox}[title=Router Prompt (condensed), fontupper=\footnotesize, colback=gray!5, colframe=gray!50]
% \textbf{System:} Given a user query and a brief page context, classify the query into one of two features: FIND (factual lookup), or GUIDE (step-by-step task). Return the mode label and a one-sentence justification.\\[2pt]
% \textbf{User:} Query: \textit{``\{query\}‘’} \quad Page context: \textit{\{page\_title, content\_type\}}
% \end{tcolorbox}

\subsection{\FIND Mode: Grounded Information Retrieval}

\FIND addresses a core limitation of existing LLM-based web agents: answers that are fluent but unverifiable because there is no pointer on the screen for users to follow or locate information. Given query $q$ and element index $\mathcal{D}$, \FIND prompts the LLM (see the full prompt in \Cref{fig:find_prompt} in Appendix) to produce a natural language answer in which every factual claim is annotated with an inline citation \hotcite{N}{exact phrase/image}, where \texttt{N} is the SoM index of the supporting HTML DOM element containing \texttt{exact phrase or image evidence}. For example, a \FIND answer for \textit{``Find when the Rec Center closes on Friday''} is: \hotcite[red!40]{45}{Fri 21 Rec Center 4-9p}, where \textit{45} is the HTML DOM index of the timetable, and \textit{Fri 21 Rec Center 4-9p} is the visual evidence captured by the agent.

When users click to each citation: the agent will automatically scroll to the corresponding evidence wrapped in an animated overlay (\eg bounding boxes, arrows, or circles). The final state visible to the user has two coupled components (see \Cref{fig:three_figures_find_example}b):
\begin{itemize}
    \item \textbf{In-page highlights:} each cited evidence is highlighted directly on the page; clicking a highlight re-scrolls and re-pulses the element.
    \item \textbf{Side panel answer:} the full answer with all citations rendered as clickable anchors is shown in a collapsible panel alongside the page.
\end{itemize}

This keeps evidence \emph{in situ}: the user verifies a claim by glancing at the highlighted element on the same page rather than switching context.

\noindent\textbf{Find Prompt (condensed)}
\begin{framed}
\textbf{System:} Given the page content and a page index (one line per element: \texttt{[N] (role) text/image}), answer the question in natural language. For every factual claim, insert an inline citation \texttt{[N:“exact phrase/image region”]}, where \texttt{N} is the index of the supporting element and \texttt{exact phrase/image region} is a short, verbatim span or a region on an image.\\[2pt]
\textbf{User:} Query: \textit{``\{query\}‘’} \quad Page content: \textit{\{pageContent\}} \quad Page index: \textit{\{[N] (role) text, image region, etc\}}
\end{framed}

% \begin{tcolorbox}[title=Find Prompt (condensed), fontupper=\footnotesize, colback=yellow!10, colframe=yellow!60!black, coltitle=black, colbacktitle=yellow!40]
% \textbf{System:} Given the page content and a page index (one line per element: \texttt{[N] (role) text}), answer the question in natural language. For every factual claim, insert an inline citation \texttt{[N:“exact phrase”]}, where \texttt{N} is the index of the supporting element and \texttt{exact phrase} is a short, verbatim span from it. If the answer is not on the page, say so explicitly and answer from general knowledge instead, citing a linked source.\\[2pt]
% \textbf{User:} Query: \textit{``\{query\}‘’} \quad Page content: \textit{\{pageContent\}} \quad Page index: \textit{\{[N] (role) text, ...\}}
% \end{tcolorbox}

\subsection{\GUIDE Mode: Mixed-Initiative Step-by-Step Navigation}

% \GUIDE targets procedural tasks where users need step-by-step assistance rather than a fully automated solution. Given query $q$ and element index $\mathcal{D}$, \GUIDE prompts the LLM (see the full prompt in \Cref{fig:guide_prompt} in the Appendix) to produce an ordered action plan $P = (a_1, a_2, \ldots, a_k)$, where each step $a_i$ specifies a natural language instruction, the SoM index of the target HTML DOM element, and an action type (\texttt{click}, \texttt{type}, \texttt{scroll}, or \texttt{navigate}). The \texttt{navigate} action handles cross-site tasks by loading a new URL and automatically resuming the plan on the updated page.

\GUIDE support verification in multi-step navigation tasks. Given query $q$ and element index $\mathcal{D}$, \GUIDE prompts the LLM (see the full prompt in \Cref{fig:guide_prompt} in the Appendix) to produce one step $a_i$ at a time, forming an action sequence $P = (a_1, a_2, \ldots, a_k)$ where step $a_i$ specifies a natural language instruction, the SoM index of the target HTML DOM element, and an action type (\eg \texttt{click}, \texttt{type}, \texttt{drag}, \texttt{scroll}, or \texttt{navigate}). The \texttt{navigate} action handles cross-site tasks by loading a new URL and automatically resuming the plan on the updated page. Any step may additionally attach one or more \textit{evidence} items that ground the instruction or, on the final step, the answer described in \textit{Textual and Visual Evidence} below.

\subsec{Step-by-Step Delivery.}
Steps are surfaced one at a time. When step $a_i$ is presented, the target element is highlighted with a pulsing beacon on the page, and the instruction appears as a tooltip anchored to that element. 
% The side panel displays the instruction alongside two control options (see \Cref{fig:guide_example}):
% \begin{itemize}
%     \item \textbf{Next} --- confirms the step, triggers a HTML DOM re-read, and advances to $a_{i+1}$.
%     \item \textbf{Stop} --- ends the session, leaving completed steps in place.
% \end{itemize}

For example, given the query \textit{``Find me a parking lot near ABC Hall''}, the second step highlights the button Search in a red box, and the fourth step annotates the target hall and parking lot in green and yellow box.

% \medskip
% \noindent\webhighlight[yellow!45]{Step 1:} \textit{Click} \hotcite[cyan!40]{14}{Settings} \textit{in the top-right corner of your profile.} \quad \guideboxgreen{Next} \quad \guideboxred{Stop}
% \medskip

% \noindent where \hotcite[cyan!40]{14}{Settings} simultaneously pulses on the page so the user can visually confirm the target before proceeding. This confirmation gate is the Feedback loop shown in \Cref{fig:overview}: the user retains full veto power over every action.

\noindent\textbf{Guide Prompt (condensed)}
\begin{framed}
\textbf{System:} Given a user task and a page index, produce one step at a time as JSON: a natural language instruction, the target element index, and an action type (\texttt{click}, \texttt{type}, \texttt{drag}, \texttt{scroll}, or \texttt{navigate}). When the instruction or final answer needs supporting evidence, attach evidence items, each anchored either to an element index or, if no single element captures it, to a screenshot region flagged for annotation with a short instruction. On the final step, return the answer with \texttt{[ev:key]} citations to saved evidence.\\[2pt]
\textbf{User:} Task: \textit{``\{query\}‘’} \quad Page index: \textit{\{[N] text, image region, etc\}}
\end{framed}

% \begin{tcolorbox}[
% title=Guide Prompt (condensed),
% fontupper=\footnotesize,
% colback=blue!10,
% colframe=blue!70,
% coltitle=white,
% colbacktitle=blue!80
% ]
% \textbf{System:} Given a user task and a page index, produce one step at a time as JSON: a natural language instruction, the target element index, and an action type (\texttt{click}, \texttt{type}, \texttt{drag}, \texttt{scroll}, or \texttt{navigate}). When the instruction or final answer needs supporting evidence, attach evidence items, each anchored either to an element index or, if no single element captures it, to a screenshot region flagged for annotation with a short instruction. On the final step, return the answer with \texttt{[ev:key]} citations to saved evidence.\\[2pt]
% \textbf{User:} Task: \textit{``\{query\}‘’} \quad Page index: \textit{\{[N] text, ...\}}
% \end{tcolorbox}

\subsec{Textual and Visual Evidence.}
To keep \GUIDE's instructions and final answers as verifiable as \FIND's, any step may attach one or more evidence items justifying a claim. \agent grounds each item in one of two ways, chosen by the LLM depending on whether the evidence corresponds to a single page element:
\begin{itemize}
    \item \textbf{Textual evidence:} when the evidence is an indexed text span, or button, the item is anchored to its SoM index and reuses \FIND's highlight-and-scroll mechanism.
    \item \textbf{Visual evidence:} when the evidence is images, charts, or graphs or the relationship between objects (\eg a spatial relationship such as ``the parking lot is next to ABC Hall''), the item is flagged for annotation with a short natural-language instruction. A dedicated \textbf{evidence annotator} call then receives the current screenshot and this instruction, and returns a small set of normalized shapes (\eg box, ellipse, arrow, or line) that are drawn directly onto a cropped screenshot (see \Cref{fig:three_figures_find_example}b).
\end{itemize}
Both evidence kinds are referenced from the step instruction or final answer with an inline citation \texttt{[ev:key]}; clicking a citation reveals the corresponding DOM highlight or annotated screenshot inline in the answer. For the parking-lot example in \Cref{fig:guide_example}, the final step returns an answer such as \textit{``The nearest lot [ev:parking\_next\_to\_hall] is directly across from ABC Hall,''} where the evidence annotator boxes both locations and draws a labeled arrow between them.

\noindent\textbf{Evidence Annotator Prompt (condensed)}
\begin{framed}
\textbf{System:} You are a screenshot evidence annotator. Given the current screenshot and a short instruction describing what to show, return up to 5 normalized (0--1) shape annotations---\texttt{box}, \texttt{ellipse}, \texttt{arrow}, or \texttt{line}---each with a short label; use boxes/ellipses for objects and arrows/lines for relationships or direction.\\[2pt]
\textbf{User:} Annotation instruction: \textit{``\{annotation\_prompt\}‘’} \quad Screenshot: \textit{\{current viewport image\}}
\end{framed}

\section{Model Performance and Annotation Reliability}
\label{sec:model_performance}
Before deploying the tasks in the user study, we evaluate the underlying model outputs to characterize task difficulty and verify the quality of the annotations used to determine answer correctness and supporting evidence.

\begin{table*}[t]
\centering
\caption{Gemini~3 Flash performance on the \FIND and \GUIDE task sets. Evidence precision, recall, and F1 measure agreement between model-identified evidence and the annotated supporting evidence.}
\Description{A table compares model answer accuracy and evidence quality for FIND and GUIDE. Columns report answer accuracy, evidence precision, evidence recall, and evidence F1 score, in that order. FIND values are 0.70, 0.70, 0.60, and 0.65. GUIDE values are 0.21, 0.143, 0.194, and 0.165. F1 is the harmonic mean of precision and recall.
}
\label{tab:model_performance}
\begin{tabular}{lcccc}
\toprule
\textbf{Task} & \textbf{Answer Acc.} & \textbf{Evidence Precision} & \textbf{Evidence Recall} & \textbf{Evidence F1} \\
\midrule
\FIND  & 0.70 & 0.70  & 0.60  & 0.65  \\
\GUIDE & 0.67 & 0.47 & 0.53 & 0.49 \\
\bottomrule
\end{tabular}
\end{table*}

\agent achieves final-answer accuracies of 0.70 on \FIND and 0.67 on \GUIDE, with evidence F1 scores of 0.65 and 0.49, respectively (see \Cref{tab:model_performance}). These results indicate that the selected tasks provide sufficient difficulty for evaluating user verification in our study. To assess annotation reliability, two annotators independently label the correctness of the model's final answers and the supporting evidence for both \FIND and \GUIDE tasks. The annotators achieve substantial agreement for both task types, with Cohen's $\kappa=0.84$ for \FIND and $\kappa=0.65$ for \GUIDE. The annotators discussed any conflicting judgments until they reached consensus on the final labels.

\section{User Study}
\subsection{Data Collection}
\label{sec:data_collection}
We evaluate users' ability to verify agent responses in two settings: \FIND and \GUIDE. In ten \FIND tasks, users locate evidence supporting an agent's answer; in twelve \GUIDE tasks, they verify supporting evidence in each navigation step. Gemini~2.5 Flash~\citep{comanici2025gemini25pushingfrontier}, and Gemini~3 Flash~\citep{google2025gemini3flash} are used to generate the correct/incorrect responses presented in the user study. \Apref{sec:data_generation_and_stats} details this comparison and the dataset-construction process.

\subsec{\FIND Tasks.} The \FIND tasks focus on long, information-dense webpages, such as transit timetables, maps, and Google Calendar views, where supporting evidence may be distributed across text, images, canvas and page layouts. Natural Questions~\citep{naturalquestions2019} is a question-answering benchmark built from real Google search queries paired with short Wikipedia pages. While it evaluates whether a model can recover the correct answer from a webpage, it does not explicitly evaluate how users seek out and verify information within real, cluttered, long webpages. In contrast, WebArena and VisualWebArena~\citep{webarena,visualwebarena} are designed to evaluate autonomous web agents on multi-step interaction tasks, such as navigating between pages, clicking interface elements, and completing actions. Therefore, those benchmarks primarily focus on navigation tasks rather than in-page information seeking. To address this gap, we construct five \FIND-Text and five \FIND-Visual tasks, each requiring users to verify two evidence units. In \FIND-Text, both units are available as text; in \FIND-Visual, at least one must be inferred from an image or from spatial relationships among image objects. For each task, we create one correct and one incorrect agent response, yielding 20 responses in total for 10 tasks.

\subsec{\GUIDE Tasks.} The \GUIDE tasks examine verification during multi-step web navigation. We use Online-Mind2Web~\citep{onlinemind2web} as the task source because it contains tasks from real, common websites. VisualWebArena~\citep{visualwebarena} also provides visually grounded tasks, but its three synthetic environments (Shopping, Reddit, and Classifieds) cover fewer domains and may differ from websites familiar to participants. Six Online-Mind2Web tasks are retained as \GUIDE-Text tasks, and six other tasks are adapted into \GUIDE-Visual tasks. For each \GUIDE task, the composite query requires the agent to first extract information (text or visual content) from an intermediate step and then use that information to complete a subsequent step (\eg \textit{Find a photo related to Trinity in The Matrix Wiki community and add the clothes she is wearing to Amazon cart}). Each \GUIDE task type contains six variations (three with correct agent responses and three with incorrect agent responses), for a total of 12 tasks. 

\subsection{Study Design}
\begin{figure*}[t]
    \centering
    \includegraphics[width=\textwidth]
        {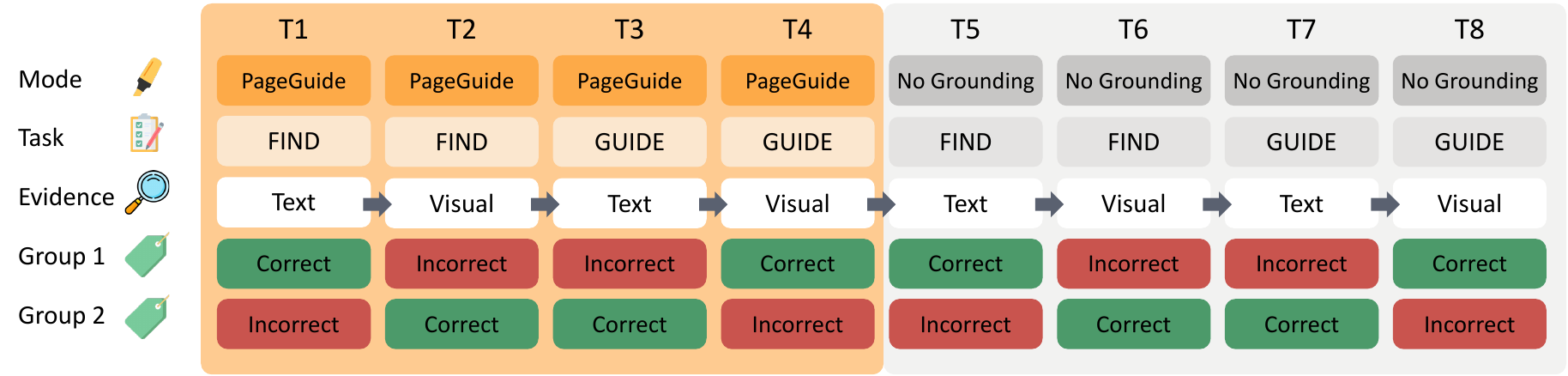}
    \caption{\textbf{User study design.} Each participant completes eight tasks spanning two grounding conditions (PageGuide vs.\ no grounding), two interaction modes (\FIND and \GUIDE), and two evidence types (text and visual). Participants are assigned to one of two counterbalancing groups that reverse the correct/incorrect answer assignments across tasks, ensuring that each experimental setting is represented equally across the study.}
    \Description{A study-protocol diagram arranges eight tasks, T1 through T8, from left to right. The first four tasks use PageGuide and the last four use no grounding. Each block combines FIND and GUIDE with text and visual evidence. Two group rows reverse the correct and incorrect answer assignments for every task.}
    \label{fig:user_study_display}
\end{figure*}

We conduct a controlled user study (N=\users) using Gemini-3.0-Flash \cite{google2025gemini3flash}, with the temperature of 0.7 as the backbone of \agent to evaluate whether \agent improves users' ability to verify the intermediate reasoning (steps) and final answers (final states) of a specific task. 

\looseness=-1\noindent\textbf{Participants and Setup.} We recruit participants through departmental mailing lists and laboratory networks. All participants are students, ranging from undergraduate to graduate level. The study (see the interface of the study in \Cref{fig:study_display_tasks} in Appendix) is conducted in person in a controlled lab setting and last approximately 30 minutes per session. 

Participants complete eight tasks under two conditions: (1) a non-grounded condition, where they used \agent with highlights and references turned off and (2) a grounded condition, where \agent is enabled (see \Cref{fig:user_study_display}). Each task lasts up to three minutes, and the tasks span all combinations of two grounding conditions (non-grounded vs. PageGuide), two interaction modes (\FIND vs. \GUIDE), and two evidence types:  \textbf{text evidence}, where the supporting information appears as HTML text on the webpage, and \textbf{visual evidence}, where the supporting information is embedded in visually rendered content such as maps, images, or charts. To balance answer correctness, participants are assigned to one of two counterbalancing groups with opposite correct/incorrect labels. Each participant also does not see the same task twice. All procedures were reviewed and approved by the Institutional Review Board (IRB).

\looseness=-1 For each task, we record the verification time, accuracy, and manual behavioral signals such as Scrolling, Clicking, and Text Selection. Participants also completed post-task surveys to report their subjective perceptions of the system. The survey questions are designed in a Likert-scale and shown in \Cref{fig:post_survey}.

\subsection{Quantitative Results}
\label{sec:quantitative_results}
Next, we present experimental results from our user study and address the three research questions introduced in \Cref{sec:introduction}. \\

\begin{figure*}[t]
    \centering
    \includegraphics[width=1\textwidth]{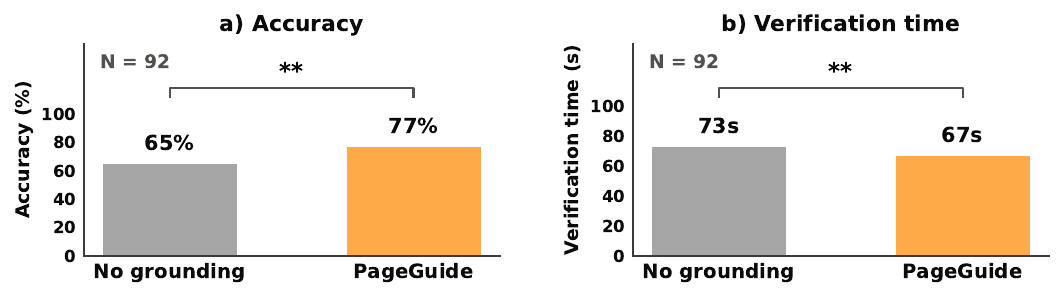}
    \caption{\textbf{Effect of grounded evidence (\agent) on task performance.} Across \FIND and \GUIDE tasks ($N=92$ users), \agent improves accuracy from 65\% to 77\% (a) and reduces verification time from 73s to 67s (b). ($^{**}p<0.01$)}
    \label{fig:acc_and_time}
    \Description{Two bar charts compare verification accuracy and time with no grounding and PageGuide. The horizontal axes identify the conditions. The left vertical axis shows accuracy in percent: 65 for no grounding and 77 for PageGuide. The right vertical axis shows verification time in seconds: 73 and 67, respectively. Brackets mark both comparisons as statistically significant, with p less than 0.01.}

\end{figure*}

\subsec{\textbf{RQ1) Grounded Evidence Yields Consistent Accuracy Gains across Tasks and Evidence Types}}
\label{sec:result2}

Across all \FIND and \GUIDE trials \modeicon, \agent improves task accuracy from 65\% in the non-grounded condition to 77\% (\increasenoparent{12 pp}) in \Cref{fig:acc_and_time}. Because each participant completes multiple trials under both conditions, we analyze trial-level accuracy using a logistic mixed-effects model, accounting for repeated observations across participants and items. The model estimates a positive effect of \agent on accuracy ($\beta=0.602$, $SE=0.23$, $z=2.62$, $p<0.01$). \textbf{Together, these results show that grounded evidence consistently improves users' ability to reach correct outcomes.}

This accuracy benefit holds across both task modes \taskicon. For \FIND tasks, where we deliberately use long, information-dense pages to challenge users' ability to locate and verify relevant evidence, accuracy increases from 53\% to 62\% (\increasenoparent{9 pp}). For \GUIDE tasks, accuracy increases from 76\% to 88\% (\increasenoparent{12 pp}). \textbf{Thus, \agent improves correctness for both evidence verification and procedural guidance tasks, despite their substantially different interaction demands.}

The improvement also holds regardless of whether the underlying answer is correct or incorrect \labelicon. For correct answers, verification accuracy increases from 71\% to 86\% (\increasenoparent{15 pp}), while for incorrect answers it increases from 60\% to 70\% (\increasenoparent{10 pp}). \textbf{The larger gain for correct answers suggests that grounded evidence is particularly effective at helping users confirm well-supported responses, while still improving their ability to detect incorrect ones.}

Lastly, \agent improves accuracy across both text and visual evidence \evidenceicon. For text-based evidence, accuracy increases from 67\% to 81\% (\increasenoparent{14 pp}); for visual-based evidence (\eg images, charts), it increases from 63\% to 74\% (\increasenoparent{11 pp}). \textbf{These results suggest that the accuracy benefit of grounding generalizes across different evidence representations rather than being limited to text.}

\subsec{\textbf{RQ2a) Grounded Evidence Reduces Verification Time across Tasks and Evidence Types}}

Across all \FIND and \GUIDE trials \modeicon, verification time decreases from 73s in the non-grounded condition to 67s with \agent (see \Cref{fig:acc_and_time}). A linear mixed-effects model, accounting for repeated observations across participants and items, estimates a 5.6s difference between conditions ($SE=1.8$, $t=3.11$, $p<0.01$). \textbf{Overall, grounded evidence reduces the time users need to evaluate and complete tasks.}

The efficiency benefit, however, varies across task modes \taskicon. For \FIND, completion time decreases only slightly from 77s to 75s, whereas for \GUIDE it decreases from 80s to 74s. \textbf{Thus, while grounding improves efficiency overall, its time benefit is more pronounced for \GUIDE; for \FIND, the primary benefit is improved correctness rather than speed.}

A similar difference emerges when separating trials by answer correctness \labelicon. For correct answers, verification time decreases substantially from 82s to 71s. In contrast, for incorrect answers, verification time remains nearly unchanged (76s vs.\ 77s). \textbf{This suggests that grounded evidence helps users confirm correct answers more quickly, while improving detection of incorrect answers without requiring additional verification time.}

Finally, the efficiency benefit depends on the form of evidence being inspected \evidenceicon. For text-based evidence, verification time remains unchanged at 70s across conditions. For visual or other non-text evidence, however, verification time decreases from 87s to 78s. \textbf{Grounding therefore provides a clearer efficiency benefit for visually rendered content, which may otherwise require more effort to locate and inspect manually.}

% \begin{figure*}[t]
%     \centering
%     \includegraphics[width=1\textwidth]{figures/time_mean.pdf}
%     \caption{Verification time (seconds) for the non-grounded and grounded (\agent) conditions. \agent reduces \FIND verification time (155.6s $\to$ 133.0s) but significantly increases \GUIDE judge time (70.4s $\to$ 109.8s), reflecting deliberation spent inspecting grounded evidence rather than inefficiency.}
%     \label{fig:copmletion_time}
%     \Description{Two grouped bar charts compare judge time between non-grounded (red) and grounded (blue). FIND judge time decreases with the extension (155.6s vs 133.0s). GUIDE judge time significantly increases with the extension (70.4s vs 109.8s), reflecting additional evidence-verification time rather than inefficiency.}
% \end{figure*}

User comments further support this finding. Without grounded evidence, 15\% of participants reported that they could not confirm what the agent had actually done. One noted, \textit{I can only see the output, so it's unclear if it hallucinated or not''}, while another resorted to guessing rather than verification: \textit{I guess I make the assumption that the 3rd item in the cart is the orange, even though I can't see that for certain.''}

\subsec{\textbf{RQ2b)} Grounded Evidence Reduces Manual Search and Inspection Effort}

\label{sec:result3}

\begin{figure*}[t]
    \centering
    \includegraphics[width=1\textwidth]{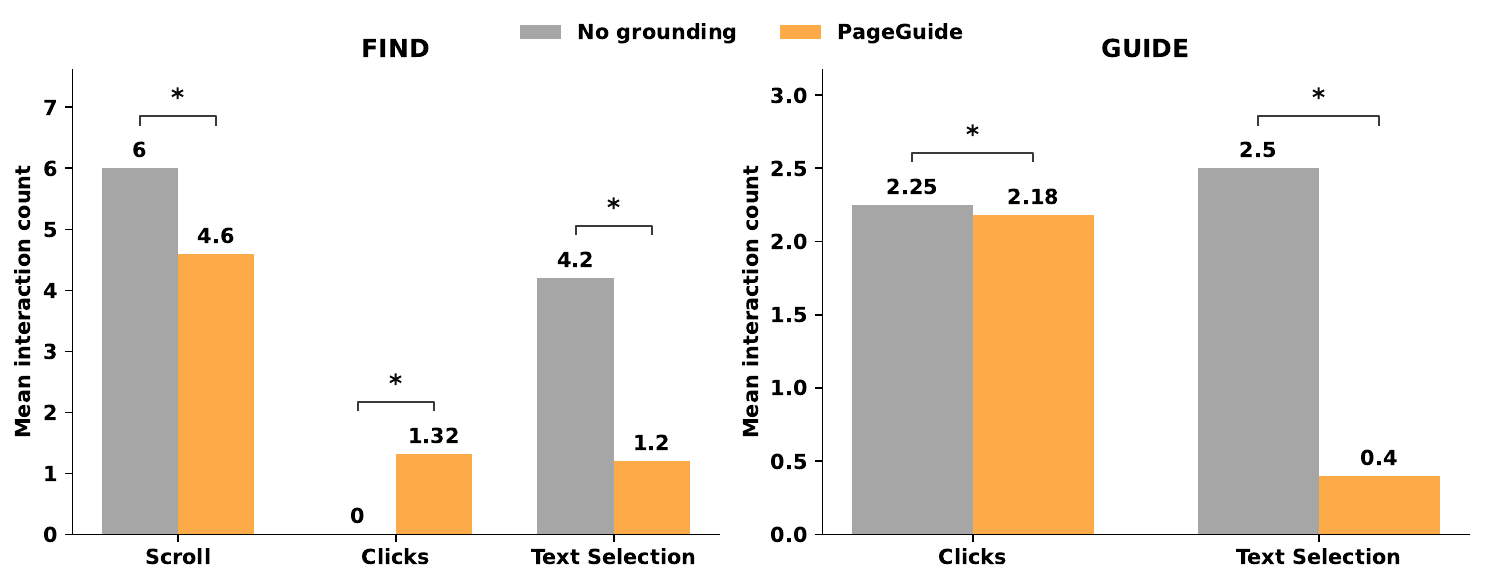}
    \caption{\textbf{Manual verification interactions across grounded (\agent) and non-grounded conditions.} Mean interaction counts for \FIND (left) and \GUIDE (right). With grounded evidence, participants perform fewer scrolls and text selections in \FIND and fewer text selections in \GUIDE. Clicks serve different purposes across conditions: grounded clicks primarily open evidence references, whereas non-grounded clicks navigate between pages for verification. Brackets indicate significant differences based on paired Wilcoxon signed-rank tests ($^{*}p<0.05$).}
    \label{fig:interaction_effort}
    \Description{Two grouped bar charts show mean verification interaction counts for FIND and GUIDE. Horizontal axes list interaction types; vertical axes show mean counts. For no grounding and PageGuide, respectively, FIND values are 6.0 and 4.6 scrolls, 0 and 1.32 clicks, and 4.2 and 1.2 text selections. GUIDE shows 2.25 and 2.18 clicks and 2.5 and 0.4 text selections; it has no scrolling category. Brackets mark each comparison with an asterisk, indicating p less than 0.05.}

\end{figure*}

Grounded evidence also reduces the manual interaction effort required for verification. Because interaction counts are discrete and non-normally distributed, we compare conditions using paired Wilcoxon signed-rank tests. In \FIND, participants scroll less with \agent (4.6 vs.\ 6.0 interactions) and perform substantially fewer text selections (1.2 vs.\ 4.2), with both differences being significant ($p<0.05$) (see \Cref{fig:interaction_effort}), suggesting that grounding reduces the need to manually search through and inspect page content. A similar pattern appears in \GUIDE, where text selection decreases from 2.5 to 0.4 interactions ($p<0.05$) (see \Cref{fig:interaction_effort}). Click counts in \GUIDE remain numerically similar (2.18 vs.\ 2.25) (see \Cref{fig:interaction_effort}), although these clicks serve different purposes: with grounding, participants primarily click evidence references, whereas without grounding, they use back and next controls to navigate between pages for verification. \textbf{Overall, the reduction in scrolling and text selection suggests that grounded evidence lowers the amount of manual searching and inspection users must perform to verify agent outputs.}

\subsection{Qualitative Results}
\label{sec:qualitative_results}
\label{sec:result4}
\begin{figure*}[t]
    \centering
    \includegraphics[width=0.9\textwidth]{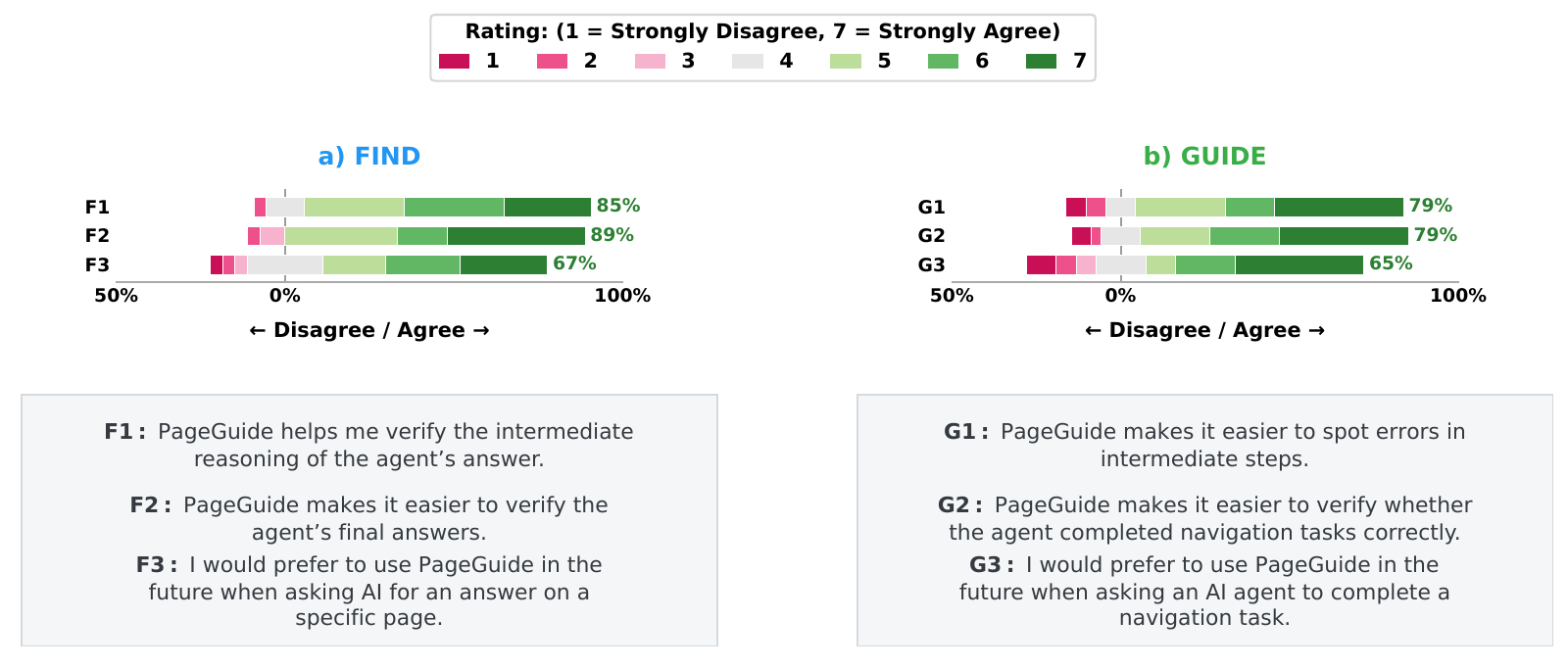}
    \caption{\textbf{Post-study perceptions of \agent for \FIND and \GUIDE.} Participants generally rated \agent positively across both modes. In \FIND, 85\% agreed that \agent helped verify intermediate reasoning, 89\% that it made the agent's final answer easier to verify, and 67\% that they would prefer to use \agent in the future. In \GUIDE, 79\% agreed that \agent made it easier to spot errors in intermediate steps, 79\% that it made task completion easier to verify, and 65\% that they would prefer to use \agent in the future. Ratings use a 7-point Likert scale from 1 (Strongly Disagree) to 7 (Strongly Agree).}
    \label{fig:post_survey}
    \Description{Two diverging stacked bar charts show seven-point agreement ratings for FIND and GUIDE. Each panel has three question rows. Horizontal segment widths represent response percentages: disagreement extends left, agreement right, and the neutral rating is centered on zero. Ratings run from 1, strongly disagree, to 7, strongly agree. Percentages beside the bars combine ratings 5 through 7: FIND has 85 percent for intermediate reasoning, 89 percent for final-answer verification, and 67 percent for future use; GUIDE has 79 percent for error spotting, 79 percent for completion verification, and 65 percent for future use. Question text appears beneath the panels.}

\end{figure*}

We collected participants' subjective feedback through a post-study questionnaire (see \Cref{fig:post_survey}) assessing perceived support for verification and preference for future use across both interaction features. The survey has six items rated from 1 (strongly disagree) to 7 (strongly agree); we report agreement as the proportion of responses rated 5--7. Using these responses, we address \textbf{RQ3} and find that \textbf{participants generally perceived \agent as helpful for verification, with more moderate support for future use}.

For \FIND tasks, 85\% of participants agreed that \agent helps them verify the intermediate reasoning of the agent's answer (F1), and 89\% agreed that it makes the agent's final answers easier to verify (F2). A smaller majority, 67\%, preferred to use \agent in the future when asking AI for an answer on a specific page (F3). For \GUIDE tasks, 79\% agreed that \agent makes errors in intermediate steps easier to spot (G1), and 79\% agreed that it helps them verify whether the agent completed navigation tasks correctly (G2). Meanwhile, 65\% preferred to use \agent in the future when asking an AI agent to complete a navigation task (G3). The positive verification ratings are consistent with the increase in overall verification accuracy from 65\% to 77\% (\Cref{fig:acc_and_time}).

Open-ended comments left after individual trials further illustrate the perceived verification benefits. Participants credited \agent's grounded evidence directly: \textit{``highlighted text helped''} and \textit{``The screenshots really helped''}. Comments in the non-grounded condition described difficulty checking the agent's outputs: \textit{``It's very difficult to double-check the agent on non-grounded tasks''} and \textit{``I don't know if the agent's answer is correct or not because I cannot see the information on the page myself.''} These comments illustrate how visible evidence can support verification, while the future-use ratings leave room to investigate factors that influence continued use.

\section{Limitations and Future Work}
\looseness=-1\subsec{Evidence highlights lack cross-page persistence.} \FIND may surface too many highlights on dense pages and currently does not preserve evidence across page visits. Future work will prioritize relevant highlights and support \textbf{multi-page context} to maintain a cross-page evidence trail.\vspace{0.03in}

\subsec{Verification-time benefits vary across tasks and answer correctness.}
Although \agent reduces overall verification time from 73s to 67s, the savings vary across task modes: \FIND verification time decreases only slightly from 77s to 75s, while \GUIDE verification time decreases from 80s to 74s. \GUIDE accuracy also increases from 76\% to 88\% (\Cref{sec:quantitative_results}). Across task modes, verification time for incorrect answers remains nearly unchanged (76s vs.\ 77s), suggesting that surfacing evidence alone does not consistently accelerate error detection. Future work will investigate whether a calibrated per-step \textbf{confidence or error-likelihood score} can help users prioritize uncertain steps. Evaluations should assess whether such cues reduce verification time while preserving users' ability to detect errors, including errors in steps presented as high confidence.

\section{Conclusion}

We presented \agent, a browser extension that grounds LLM answers directly in the HTML DOM across both interaction features: \FIND locates and highlights supporting evidence in-place so users can instantly verify answers; and \GUIDE delivers step-by-step instructions along with evidence at each step.

The user study shows that \agent improves overall verification accuracy from 65\% to 77\% and reduces verification time from 73s to 67s. For \FIND, accuracy increases from 53\% to 62\%, while verification time decreases slightly from 77s to 75s. For \GUIDE, accuracy increases from 76\% to 88\%, and verification time decreases from 80s to 74s. These findings suggest that surfacing grounded evidence supports more accurate verification across both task modes, with larger time savings for \GUIDE. Overall, \agent points toward web agents that support human verification by making their reasoning and actions easier to inspect.

\bibliographystyle{ACM-Reference-Format}
\bibliography{references}

\clearpage
\onecolumn
\appendix

% \begin{center}
% {
% \LARGE\textsc{Appendix for:}\\[0.4em]
% \LARGE\textsc{Appendix}\\[0.4em]
% \LARGE\textsc{\agent: Browser Extension to Assist Users in Navigating a Webpage and Locating Information}
% }
% \end{center}

% \vspace{2em}
% {\hypersetup{linkcolor=xwebpink}
% \tableofcontents}
% \clearpage

\section{Interface Display of the User Study}
\label{sec:user_study_interface}

\begin{figure*}[ht!]
    \centering
    \includegraphics[width=\textwidth]{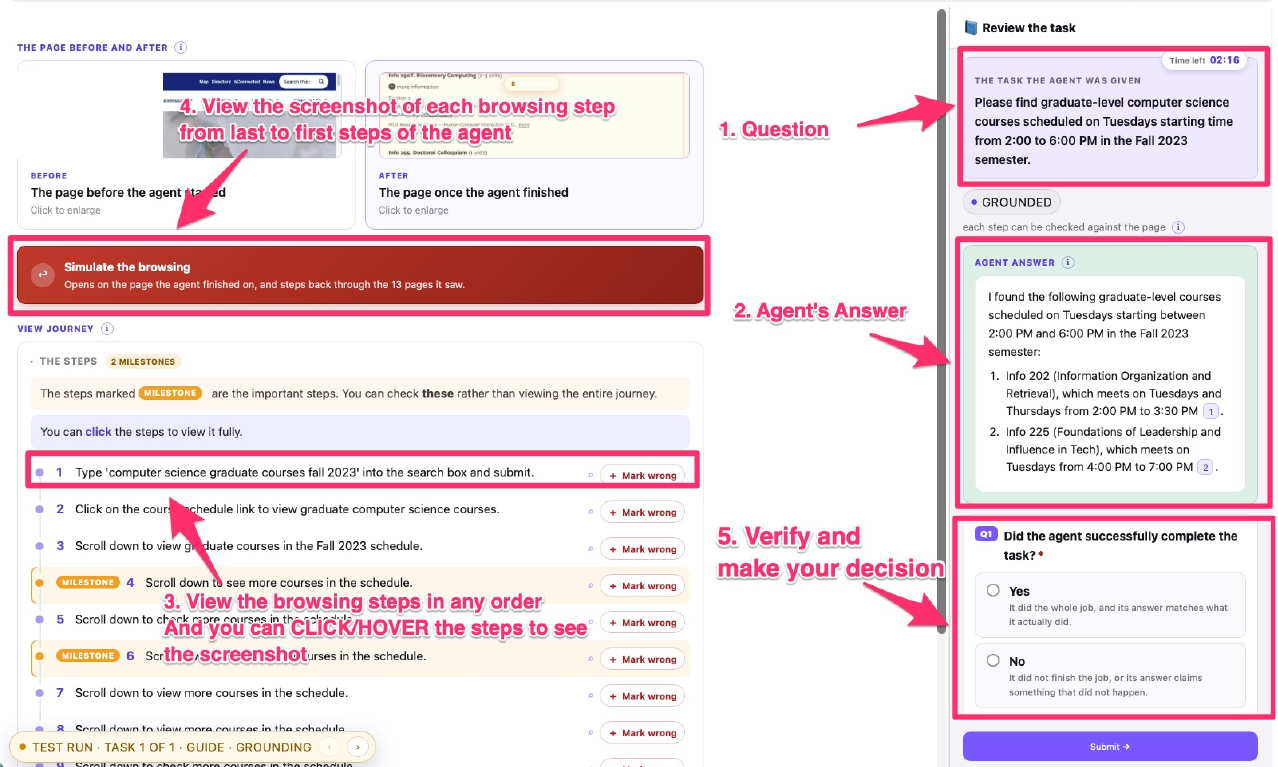}
    \caption{\GUIDE user-study interface. Participants review the task (1) and the agent's answer (2), then inspect the agent's browsing trajectory by selecting individual steps in any order (3) or replaying the full sequence. For each step, the interface displays screenshots of the webpage before and after the action (4), while milestone labels highlight steps that are particularly important for verification. After reviewing the trajectory and its supporting evidence, participants judge whether the agent completed the task successfully (5).}
    \label{fig:study_display_tasks}
    \Description{Screenshot of the GUIDE user-study interface. The right panel presents the task, the agent's answer, and a yes-or-no question asking whether the task was completed successfully. The left panel presents the agent's browsing trajectory, controls for replaying or selecting steps, milestone labels for important steps, and webpage screenshots from before and after each selected action.}
\end{figure*}

\begin{figure*}[ht!]
    \centering
    \includegraphics[width=\textwidth]{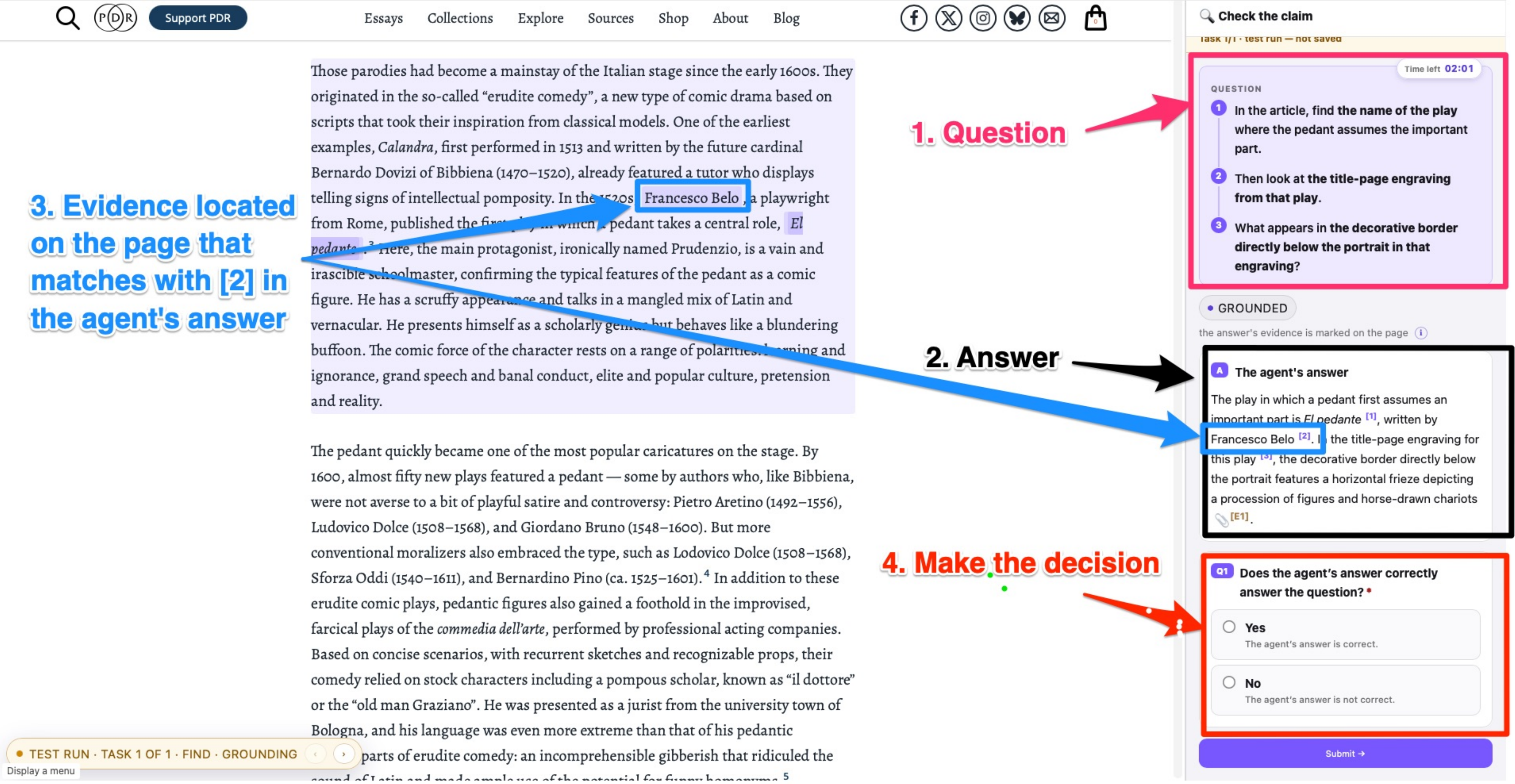}
    \caption{\FIND user-study interface. Participants review the question (1) and the agent's answer (2), then inspect the webpage to verify the answer's supporting evidence. In the grounded condition shown here, numbered references in the answer correspond to highlighted evidence on the webpage (3). After comparing the answer with the cited evidence, participants judge whether the agent answered the question correctly (4).}
    \label{fig:find_study_display}
    \Description{Screenshot of the FIND user-study interface. The webpage appears on the left, with a relevant passage highlighted as supporting evidence. The right panel presents the question, a grounded agent answer containing numbered evidence references, and a yes-or-no question asking whether the answer is correct.}
\end{figure*}

Participants evaluate outputs from either a non-grounded agent or \agent through a consistent study interface. Each task presents the task instructions, a timer, the agent's response, and a standardized judgment form. For \FIND tasks (Figure~\ref{fig:find_study_display}), participants inspect the source webpage and determine whether its content supports the agent's answer; in the \agent condition, numbered references connect the answer to highlighted evidence on the page. For \GUIDE tasks (Figure~\ref{fig:study_display_tasks}), participants review the agent's browsing steps and the corresponding before-and-after screenshots before judging whether the task was completed successfully. The consistent layout and response format enable comparable judgments across task types and study conditions.

% \begin{figure*}[tp!]
%     \centering
%     \begin{subfigure}[t]{0.48\textwidth}
%         \centering
%         \includegraphics[width=\textwidth]{figures/userstudy_display/quick_questions.pdf}
%         \caption{Quick post-task questions collecting confidence ratings and (for extension users) helpfulness ratings after each task.}
%         \label{fig:study_display_questions}
%     \end{subfigure}
%     \hfill
%     \begin{subfigure}[t]{0.48\textwidth}
%         \centering
%         \includegraphics[width=\textwidth]{figures/userstudy_display/final_state.pdf}
%         \caption{Final state of the study interface showing recorded outcomes and extension interaction history for the session.}
%         \label{fig:study_display_final}
%     \end{subfigure}
%     \caption{Post-task questionnaire (a) and session final state (b) displayed to participants during the user study.}
%     \label{fig:study_display_misc}
% \end{figure*}

\clearpage
\section{Questionnaire Display}
\label{sec:questionnaire_display}

\begin{figure*}[ht]
    \centering
    \includegraphics[width=\textwidth]{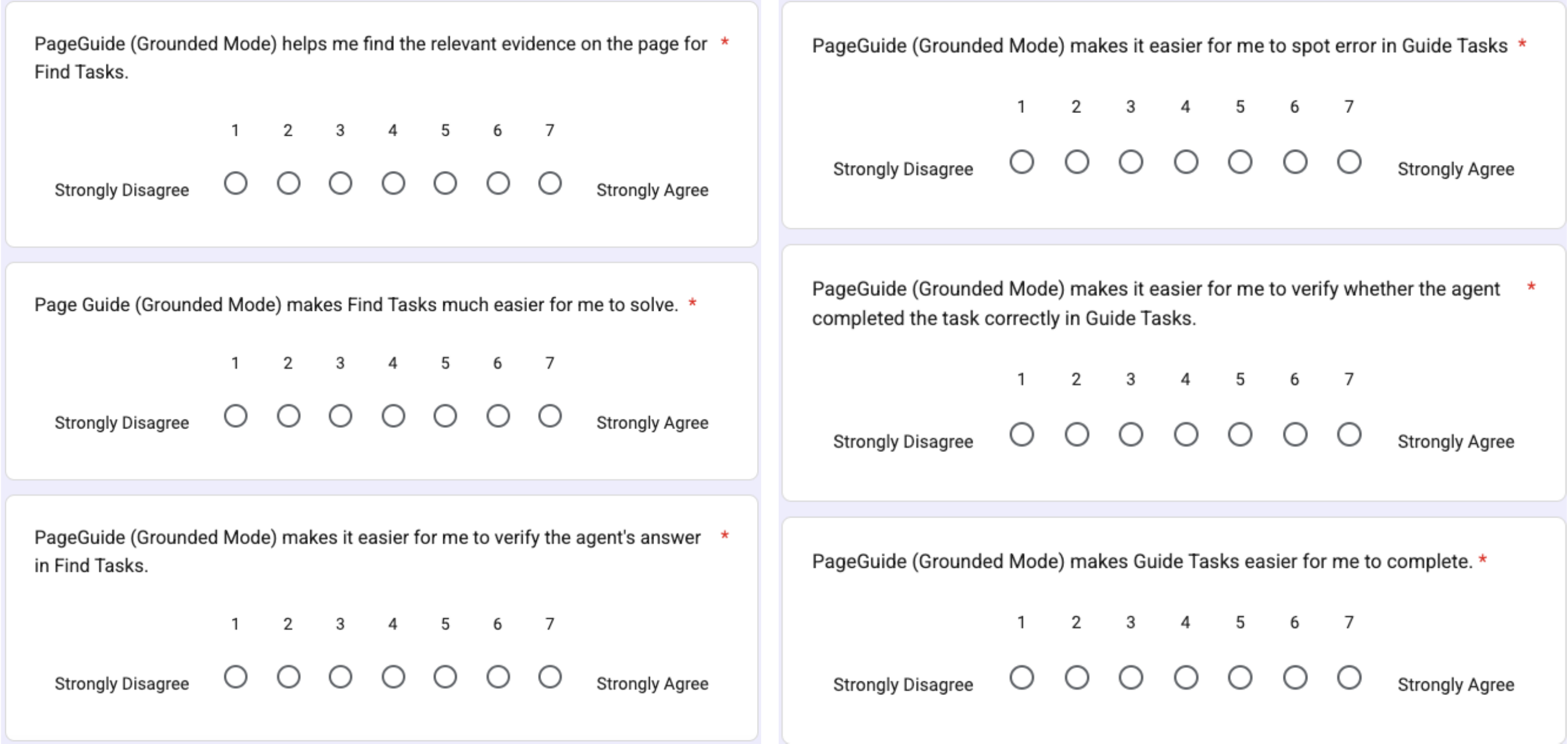}
    \caption{Post-Study Questionnaire. Find tasks require users to locate specific information on a webpage and answer correctly. In Grounded Mode, the agent provides evidence pointing to the relevant page content; in Non-Grounded Mode, it provides answers without supporting evidence. Guide tasks require users to identify errors in the agent’s trajectory. In Grounded Mode, each step includes a screenshot and visual evidence; in Non-Grounded Mode, these are omitted.}
    \label{fig:appendix_post_survey}
    \Description{Screenshot of the post-study questionnaire. It asks users about correctness, ease of use, and task difficulty across FIND, and GUIDE.}
\end{figure*}

After the experiment, we collect participants' overall feedback on their experience using the tools through a post-study questionnaire. The questionnaires are designed to capture participants' subjective impressions of the system after completing the study tasks. The full questionnaire is shown in \Cref{fig:appendix_post_survey}.

\clearpage
\section{Prompts}
\label{sec:prompts_appendix}

This section reproduces the full system prompts used by \agent's four LLM-backed components. Each prompt is designed for a specific role in the pipeline and enforces a structured output format that downstream code can parse and act on directly.

\begin{itemize}
    \item \textbf{Router Prompt} (\Cref{fig:routing_prompt}): classifies the user's natural language query into one of 2 handlers (\texttt{find}, \texttt{guide}) and returns a JSON object with a confidence score and a one-sentence justification.
    \item \textbf{Find Prompt} (\Cref{fig:find_prompt}): instructs the LLM to answer the user's question using \texttt{[N:"text/image region"]} inline citations that reference specific elements from the SoM page index, enabling the frontend to resolve and highlight each cited span.
    \item \textbf{Guide Prompt} (\Cref{fig:guide_prompt}): produces one step at a time as a JSON action object---instruction text, target SoM index, action type, and a next-step hint.
\end{itemize}

\begin{figure*}[ht!]
\centering

\begin{minipage}{0.95\textwidth}
\begin{framed}
You are a query router for a web assistant. Your job is to classify the user's query and route it to the appropriate handler.

AVAILABLE HANDLERS:

1. "guide" - For step-by-step "how to" questions that need interactive guidance

2. "find" - For questions, information lookup, finding content, highlighting elements (DEFAULT)

ROUTING RULES:
- "guide": User wants to LEARN how to do something in steps (\eg "how do I report this video?", "where can I find settings?", "help me delete my account")

- "find": Questions about the page, finding information, showing/highlighting elements (\eg "what is this page about?", "find the price", "show me images", "where is the login button?")

Return JSON only:

{

  "handler": "guide" | "find",
  
  "confidence": 0.0-1.0,
  
  "reason": "Brief explanation of why this handler"
  
}

EXAMPLES:

Query: "What is the price of this product?"

→ {"handler": "find", "confidence": 0.9, "reason": "Question about page content"}

Query: "How do I report this video?"

→ {"handler": "guide", "confidence": 0.9, "reason": "How-to question needing step-by-step guidance"}

Query: "Where can I buy the item in my image?"

→ {"handler": "image\_find", "confidence": 0.95, "reason": "Question about uploaded image, finding on page"}

Query: "Find where it mentions the methodology"

→ {"handler": "pdf\_find", "confidence": 0.85, "reason": "Finding specific content in a document"}
\end{framed}
\end{minipage}
\caption{Routing prompt used to classify user queries into appropriate handlers (\eg \texttt{find}, \texttt{guide}.}
\label{fig:routing_prompt}
\Description{Text box showing the full Router prompt. The system message instructs the LLM to classify a user query into FIND, or GUIDE using canonical examples, and return a mode label plus a one-sentence justification. The user message template includes the query and a compact page context (title, content type).}
\end{figure*}

\begin{figure*}[t]
\centering

\begin{minipage}{0.95\textwidth}
\begin{framed}
You are a helpful web assistant. Answer the user's question based on the page content, using inline citations.

PAGE CONTENT:
\{pageContent\}

PAGE INDEX (use these numbers for citations):
\{pageIndex\}

INSTRUCTIONS:

1. Answer the question based on the page content if possible

2. If the page content has the answer, use [N:"text"] citations inline to reference specific elements from the PAGE INDEX

   - N is the index number from PAGE INDEX
   
   - "text" is the EXACT text snippet to highlight (copy from the page content)
   
3. Each citation should point to an element that supports that part of your answer

4. For lists of items, cite each one with the specific text to highlight

5. Use ONE citation per item (if same text has multiple indices, pick the link)

6. The "text" should be a short, specific phrase (not the entire element text)

7. Consider conversation history for context, but always answer based on CURRENT page content

8. NEVER reproduce existing footnote markers from the webpage itself (e.g. Wikipedia's [1], [2], [3]) — only use [N:"text"] format where N comes from the PAGE INDEX above

9. **CRITICAL**: If the information is NOT provided on this page:

   - State exactly: "The information is not provided on this page."
   
   - Then, providing the answer using your own general knowledge base is HIGHLY ENCOURAGED. Do not simply stop after stating it is not on the page.
   
   - You MUST include citations to real, valid source URLs using STANDARD MARKDOWN LINKS. Wrap the link in text so the user can click the hyperlink, \eg [Text to display](https://url-of-source.com).
   
   - Whenever possible, append Chrome Text Fragments 
   ('\#:\~:text=exact\%20phrase') to the URL. This allows the browser to automatically highlight the specific text when the user opens the citation.
   
   - Example when not on page: "The information is not provided on this page. However, the tallest building in the world is the [Burj Khalifa](https://en.wikipedia.org/wiki/Burj\_Khalifa\#:\~:text=tallest..)."

CITATION EXAMPLE:

Question: "Who directed this movie?"

Answer: The movie was directed by Christopher Nolan [45:"Christopher Nolan"].

Question: "Who are the main actors?"

Answer: The main actors are Leonardo DiCaprio [23:"Leonardo DiCaprio"], Tom Hardy [27:"Tom Hardy"], and Ellen Page [31:"Ellen Page"].

Answer the user's question with citations:
\end{framed}
\end{minipage}
\caption{Prompt used to answer user queries grounded in webpage content, producing responses with interleaved references to supporting evidence spans (\eg The movie was directed by Christopher Nolan [45:"Christopher Nolan"]).}
\label{fig:find_prompt}
\Description{Text box showing the full FIND mode prompt. The system message instructs the LLM to answer the query in natural language and insert inline citations in the format [N:"exact phrase"] for every factual claim, where N is the SoM element index. The user message template includes the query and the DOM element index.}
\end{figure*}

\begin{figure*}[t]
\centering

\begin{minipage}{0.95\textwidth}
\begin{framed}
You are a helpful guide assistant. Users ask "how to" questions and you provide step-by-step guidance.

You will receive:

1. PAGE INDEX - Visible elements on the page

2. USER QUESTION - What the user wants to do

3. STEP NUMBER - Current step (1 = first step)

4. PREVIOUS STEPS - What was done before (if any)

Your job: Guide the user ONE STEP at a time.

IMPORTANT CONCEPTS:

- Some buttons/options are HIDDEN in menus (like "..." or "\vdots" three-dot menus)

- If the target isn't visible, guide user to open the menu FIRST

- Common hidden locations: dropdown menus, "More" buttons, three-dot menus, right-click menus, settings icons

Return JSON:

{

  "step": 1,
  
  "instruction": "Clear instruction for this step",
  
  "highlight": {"index": N, "text": "element to highlight"},
  
  "waitFor": "click" | "input" | "scroll" | null,
  
  "isLastStep": false,
  
  "nextStepHint": "What will happen next"
  
}

RULES:

1. ONE step at a time - don't overwhelm the user

2. If target is likely hidden in a menu, first step should open that menu

3. Use "waitFor": "click" when user needs to click something

4. Set "isLastStep": true only when the goal is achieved

5. Make instructions clear and specific

6. Highlight the element user needs to interact with

EXAMPLES:

PAGE INDEX:

[5] (button) \vdots

[12] (button) Share

[15] (button) Save

Q: "How do I report this video?" (Step 1)

→ \{
"step":1,"instruction":"Click the three-dot menu (\vdots) to see more options",
"highlight":{"index":5,"text":"\vdots"}, "waitFor":"click","isLastStep":false,"nextStepHint":"The menu will open with Report option"
\}

Q: "How do I report this video?" (Step 2, after menu opened)

PAGE INDEX now shows: [20] (button) Report

→ \{"step":2,"instruction":"Now click 'Report' to report this video",

"highlight":{"index":20,"text":"Report"},"waitFor":"click","isLastStep":true,"nextStepHint":"You'll see reporting options"
\}
\end{framed}
\end{minipage}
\caption{\GUIDE mode prompt. The LLM receives the current page element index, the user query, the current step number, and any previously completed steps. It returns one action at a time as a JSON object specifying the instruction text, the SoM index and text of the target element, the action type (\texttt{click}, \texttt{input}, \texttt{scroll}), and a next-step hint.}
\label{fig:guide_prompt}
\Description{Text box showing the full GUIDE mode prompt. The system message instructs the LLM to produce an ordered action plan where each step specifies a natural language instruction, target element index, action type (click, type, scroll, or navigate), and a one-sentence hint. The user message template includes the task query and DOM index.}
\end{figure*}

\section{Model Details}

All LLM calls in \agent are served through \textbf{OpenRouter}\footnote{\url{https://openrouter.ai}}, a unified API gateway that provides access to multiple model providers under a single endpoint. We evaluate two Gemini models throughout this paper:

\begin{itemize}
    \item Gemini 2.5 Flash \texttt{google/gemini-2.5-flash-preview}
    \item Gemini 3.0 Flash \texttt{google/gemini-3.0-flash-preview}
\end{itemize}

\clearpage
\section{Dataset Construction and Characterization}
\label{sec:data_generation_and_stats}

\begin{figure*}[ht!]
    \centering
    \includegraphics[width=\textwidth]{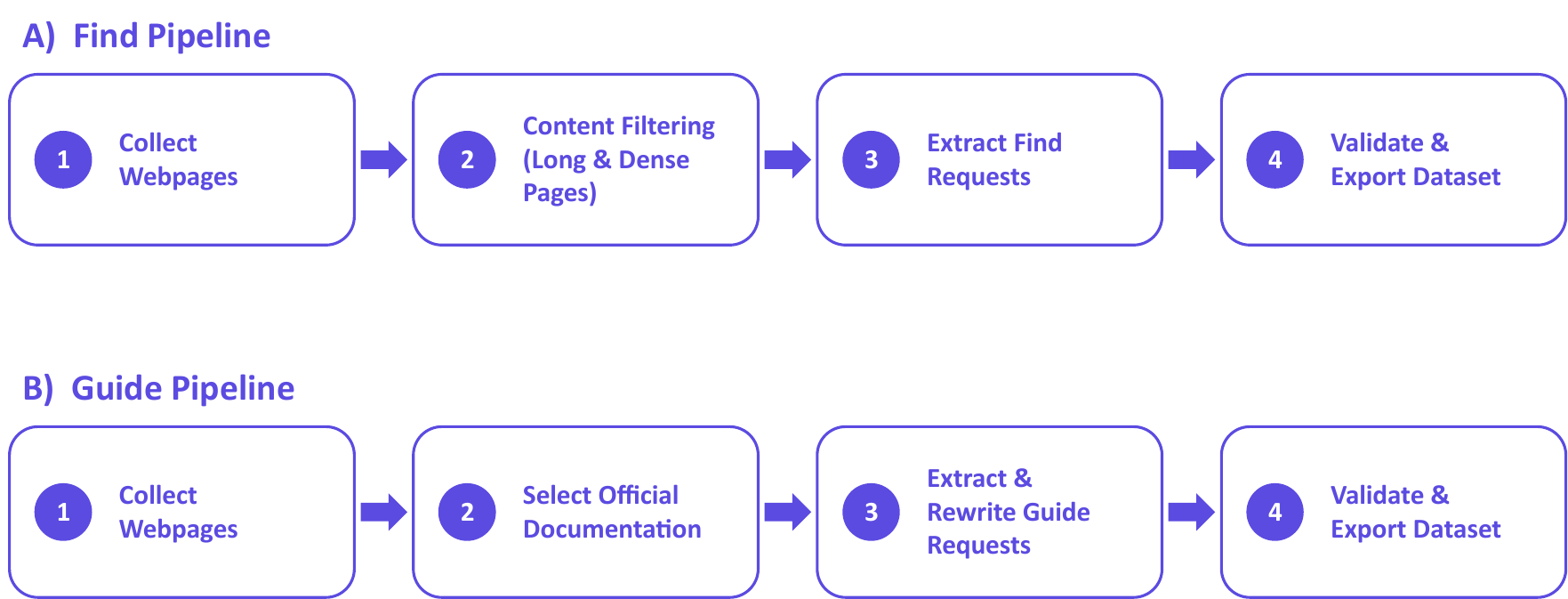}
    \caption{Overview of the data-generation pipelines for PageGuide’s two datasets. (A) For Find, we collect webpages, retain those that are long and information-dense, generate information-finding requests, and validate the resulting examples. (B) For Guide, we collect webpages, retain tasks that require multiple steps and can be completed successfully, generate and refine navigation requests, and validate the resulting examples.}
    \label{fig:appendix_data_generation_pipeline}
    \Description{Three-panel diagram of data generation pipelines. Panel A (FIND): webpages are collected, filtered for content density, and question-answer pairs are LLM-generated and verified. Panel B (GUIDE): tasks are sampled from web agent benchmarks, filtered for human-supervision relevance, and correctness assessed by terminal-state screenshots.}
\end{figure*}

Figure~\ref{fig:appendix_data_generation_pipeline} summarizes the construction pipelines for the \FIND and \GUIDE datasets. For \FIND, we first collect candidate webpages and filter them according to page length and content density so that the retained pages require meaningful search rather than simple fact lookup. We then construct requests whose answers depend on locating and combining evidence from different regions of the page. For example, a task based on the Harry Potter Wikipedia page requires participants to identify two nonadjacent pieces of information. For \GUIDE, we collect candidate web tasks and retain those containing multiple constraints, dependent subtasks, and a clearly verifiable outcome. For example, the request ``Find graduate-level computer science courses offered in Fall 2023 that meet on Tuesdays and start between 2:00 and 6:00 p.m.'' requires users to jointly consider the course level, subject, semester, meeting day, and start time. We extract and standardize the selected requests before validating each task for clarity, answerability, webpage accessibility, and the correctness of its reference answer or final state. The validated tasks are then exported in a consistent format for the user study.

Following this process, we characterize the complexity of the ten \FIND pages by comparing them with a random sample of 20 NQ source pages rendered under identical \(1020 \times 800\) viewport conditions. As shown in Table~\ref{tab:find_nq_page_stats}, \FIND pages were consistently larger and more structurally complex than the sampled NQ pages across all measured attributes. The median \FIND page contained \(2.29\times\) as many words and \(4.33\times\) as many rendered images. It also contained \(2.26\times\) as many DOM elements and \(1.72\times\) as many links. Collectively, these differences indicate that the \FIND dataset presents substantially greater textual, visual, and structural complexity than the sampled NQ pages.

\begin{table}[t]
    \centering
    \small
    \caption{Page-level complexity of the \FIND dataset compared with 20 randomly sampled NQ source pages. Values are medians, and the final column reports the ratio between the \FIND and NQ medians.}
    \label{tab:find_nq_page_stats}
    \begin{tabular}{lrrr}
        \toprule
        \textbf{Measure} &
        \textbf{\FIND} &
        \textbf{NQ} &
        \textbf{\FIND/NQ} \\
        \midrule
        Words                    & 10,041 & 4,380  & \(2.29\times\) \\
        Characters               & 65,210 & 28,068 & \(2.32\times\) \\
        Rendered images          & 26     & 6      & \(4.33\times\) \\
        DOM elements             & 3,849  & 1,705  & \(2.26\times\) \\
        Links                    & 963    & 559    & \(1.72\times\) \\
        DOM nesting depth        & 22     & 18     & \(1.22\times\) \\
        \bottomrule
    \end{tabular}
\end{table}

\clearpage
\section{Beyond Basic Functionality: Enhancing Web Extensions with PDF Reading, Visual Question Answering, and Page-Off Support}
\label{sec:additional_functionality}

\agent extends beyond the core \textsc{Find}, and \textsc{Guide} functionalities with three additional features: PDF reading, visual question answering, and Page-Off support. These features are motivated by an important limitation of conventional web extensions: user needs are often not confined to the visible content of the current webpage. In practice, users frequently interact with documents such as PDFs, compare webpage content with uploaded images, or ask questions that go beyond the active tab. Supporting these settings allows \agent to provide a more complete and flexible browsing experience, rather than restricting assistance to page-bound interactions alone.

\begin{figure*}[ht!]
    \centering
    \includegraphics[width=\textwidth]{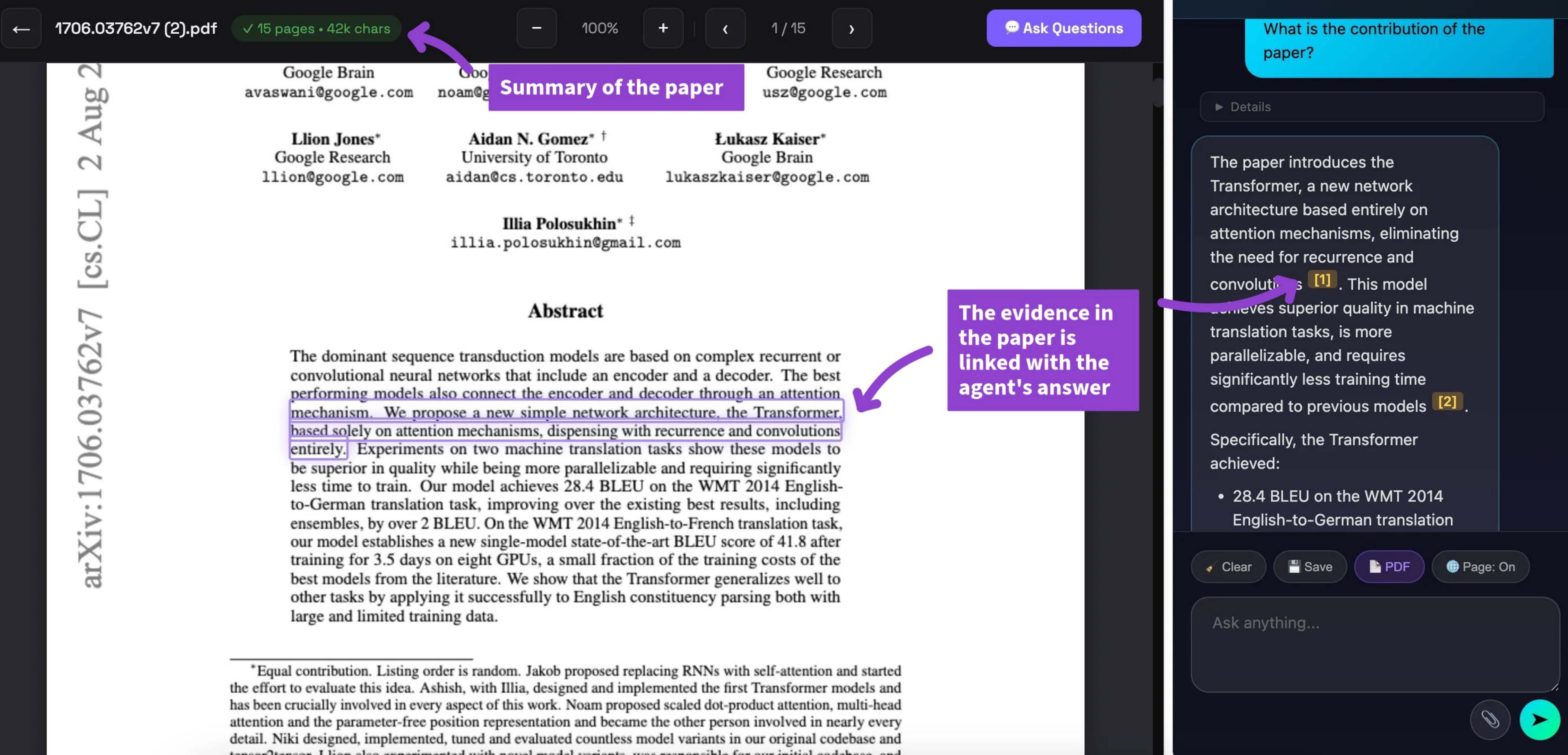}
    \caption{Users can upload a PDF and ask questions about its contents through the PDF upload feature. In this example, the user uploads the \textit{Attention Is All You Need} paper and asks about its main contributions. The agent then provides answers grounded in evidence from the document.}
    \label{fig:appendix_pdf_asking_feature}
    \Description{Screenshot of PageGuide's PDF QA feature. The user uploaded the "Attention Is All You Need" paper and asked about its main contributions. The side panel shows an answer with inline citations; cited passages are highlighted directly in the rendered PDF.}
\end{figure*}

\subsec{PDF Reading}
Many webpages contain or link to information in PDF form, such as papers, reports, manuals, and forms. However, this content is often difficult to browse efficiently using standard web tools. The PDF reading feature allows users to upload a PDF and ask questions about its content, enabling the agent to support document-centered information seeking within the same interaction framework as webpage assistance (\Cref{fig:appendix_pdf_asking_feature}).

\begin{figure*}[ht!]
    \centering
    \includegraphics[width=\textwidth]{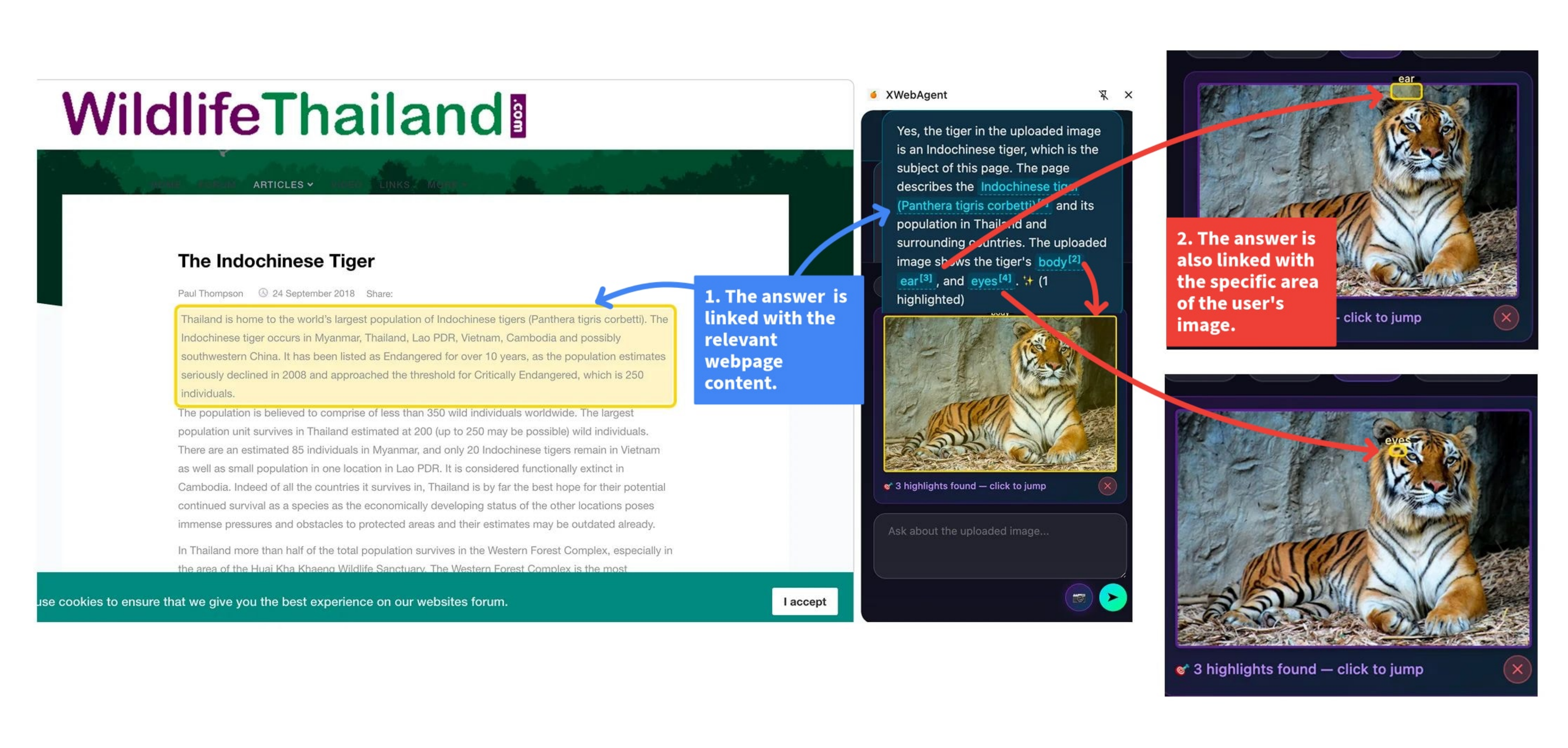}
    \caption{\agent allows users to upload an image and ask questions about it, while grounding its textual response in evidence from the webpage and specific areas of the user's image. In this example, the user asks whether the tiger in the uploaded image is the same species as the tiger described in the webpage, and the agent responds with evidence-backed information. \textit{Note:} The yellow bounding box highlighting the webpage paragraph is generated automatically by the agent.}
    \label{fig:appendix_visual_question_asking_feature}
    \Description{Screenshot of PageGuide's image QA feature. The user uploaded an image and asked a question about it. The agent provides a natural language answer grounded in webpage context, with inline citations linking claims to visible page elements.}
\end{figure*}

\subsec{Visual Question Asking}
Users also often need to reason about visual content in addition to text. For example, they may want to compare an uploaded object with items shown on a webpage, identify whether two images depict the same kind of object, or retrieve page information relevant to a visual query. The visual question asking feature allows users to upload an image and ask questions about it. The agent answers by grounding its textual response and visual highlights of the uploaded image in the textual content of the webpage, thereby extending webpage assistance to multimodal interactions (\Cref{fig:appendix_visual_question_asking_feature}).

\begin{figure*}[ht!]
    \centering
    \includegraphics[width=\textwidth]{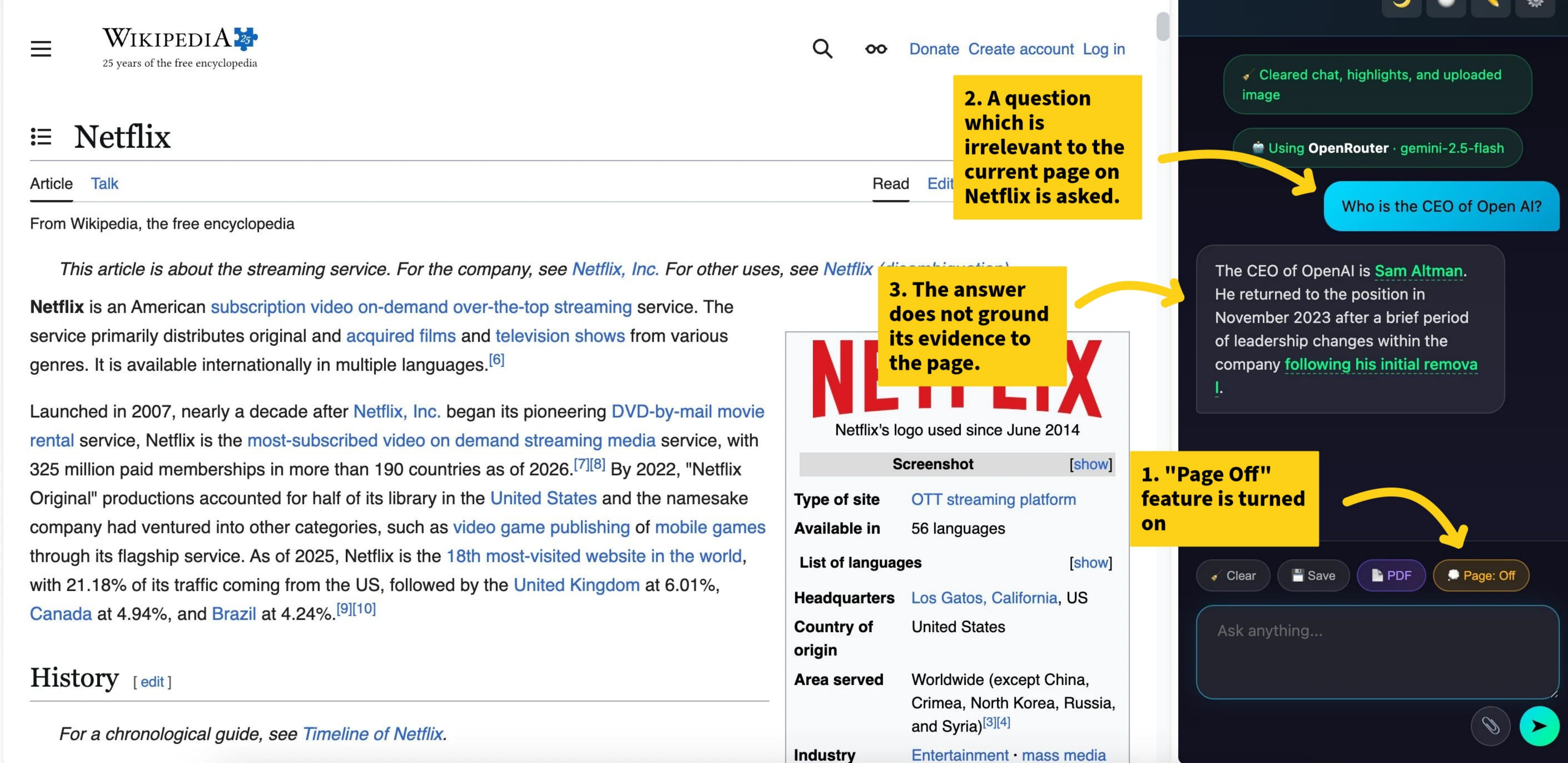}
    \caption{With the Page Off feature enabled, users can ask questions that are not related to the currently active tab. The agent answers using information gathered from other webpages on the internet, drawing on its own memory and providing supporting evidence from the internet.}
    \label{fig:appendix_page_off_feature}
    \Description{Screenshot of PageGuide's Page Off feature. In this feature, PageGuide will remove the page’s context from the input. The side panel shows a conversational answer with no in-page citations, since no page context is used.}
\end{figure*}

\subsec{Page-Off Feature}
Not all user questions are tied to the current webpage. In realistic browsing sessions, users may ask follow-up questions, seek related background information, or shift to a new intent while remaining in the extension interface. The Page-Off feature supports such free-form questions that are not directly related to the active page, allowing the extension to remain useful even when the user's information need moves beyond the current tab (\Cref{fig:appendix_page_off_feature}).

\clearpage

\section{Limitations of Existing Extensions}
\label{sec:appendix_limitations}

\begin{figure*}[ht!]
    \centering
    \includegraphics[width=\textwidth]{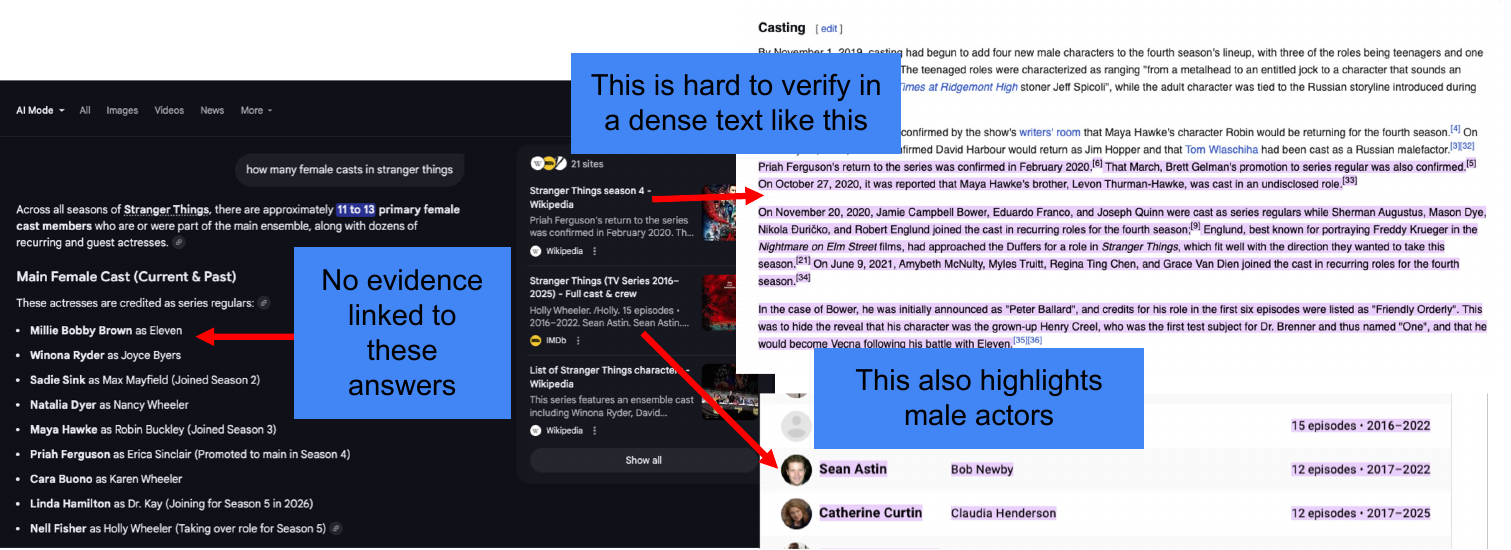}
    \caption{Example of Gemini Chatbot on the query \texttt{``How many female casts in Stranger Things?''}. The chatbot provides an answer with partial references to text spans, such as ``Stranger Things,'' but does not link evidence to individual cast names, making verification difficult. In addition, the relevant evidence is embedded within dense text sections, and the highlights include unrelated information, such as male cast members, further reducing the clarity and trustworthiness of the response.}
    \label{fig:appendix_failure_gemini_chat}
    \Description{The Gemini Extension returns an answer but users do not know where is the answer in the page. With the query "How many female casts in Stranger Things?" The chatbot returns a text answer but highlights no page elements; the user must manually verify the answer against the Wikipedia page.}
\end{figure*}

\begin{figure*}[ht!]
    \centering
    \includegraphics[width=\textwidth]{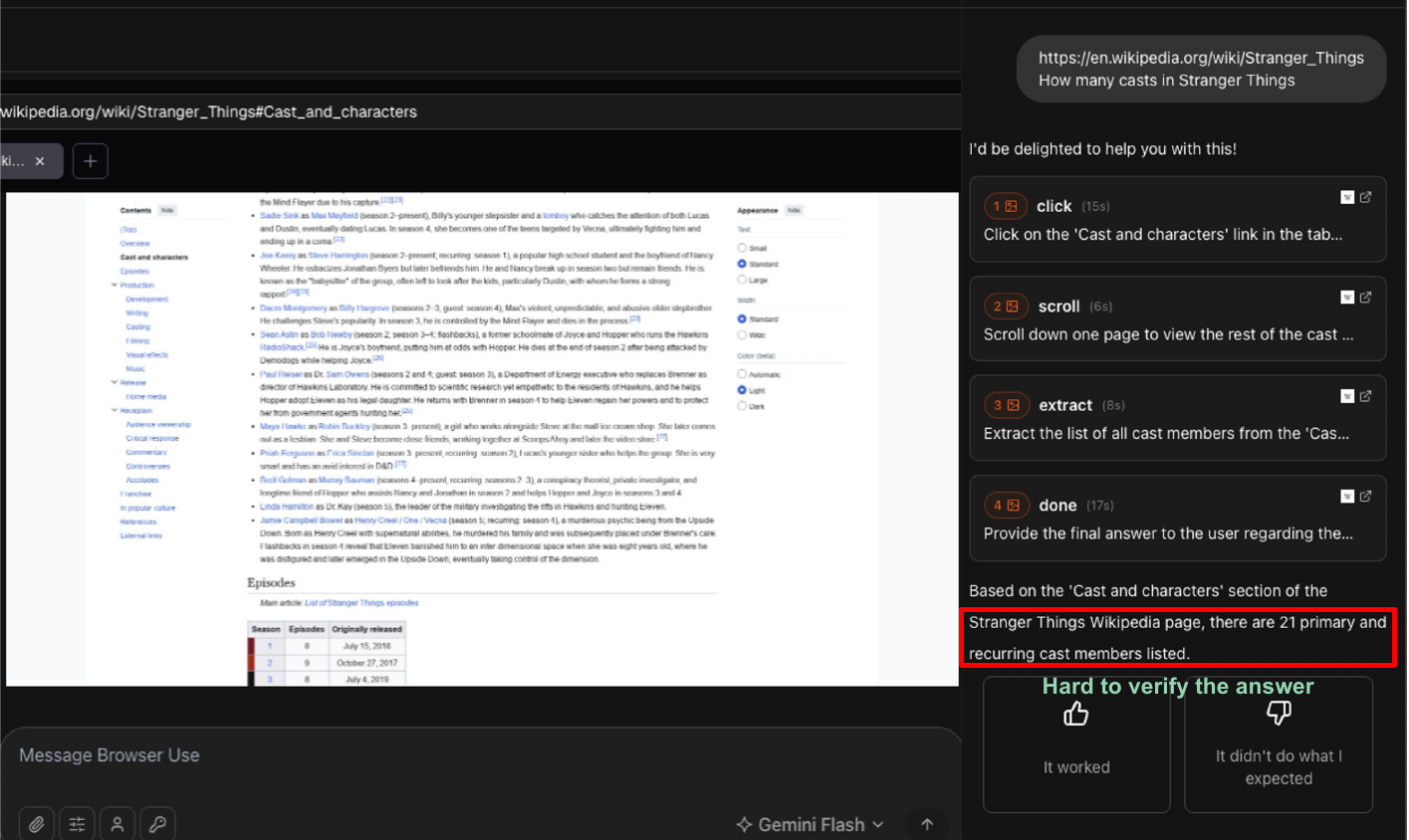}
    \caption{Example of Browser Use on \url{https://en.wikipedia.org/wiki/Stranger_Things}. The agent generates a response that correctly answers the query but is unable to highlight the corresponding evidence, such as through bounding boxes, making it difficult for users to verify the response.}
    \label{fig:appendix_failure_browser_use}
    \Description{Browser-Use returns an answer but users do not know where the answer is in the page. The agent returned a text answer in a separate panel, but the page is unmodified — no elements are highlighted and no evidence is linked.}
\end{figure*}

\begin{figure*}[ht!]
    \centering
    \includegraphics[width=\textwidth]{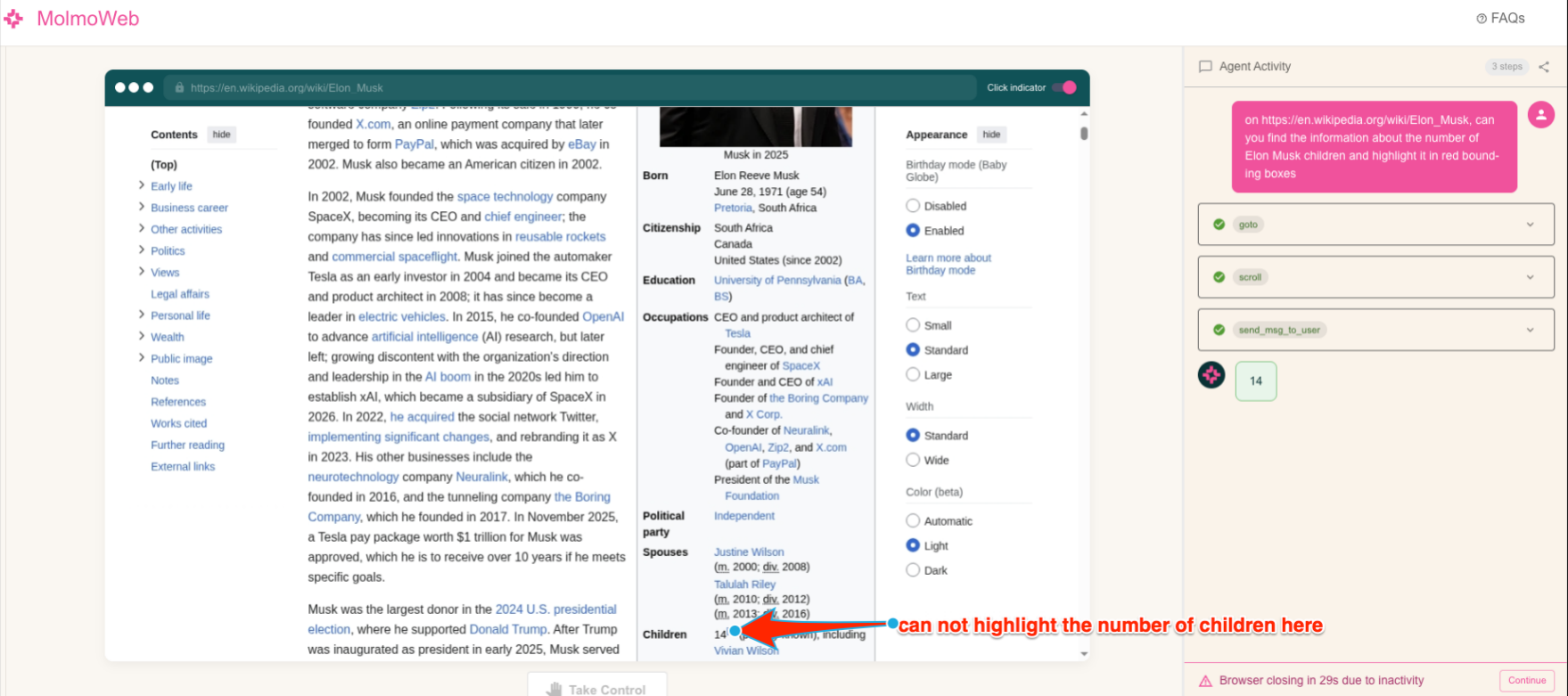}
    \caption{Example of MolmoWeb ~\cite{gupta2026molmoweb} on \url{https://en.wikipedia.org/wiki/Elon_Musk}. The agent generates a response that correctly answers the query is unable to highlight corresponding evidence, such as through bounding boxes, which makes it difficult for the users to verify the response.}
    \label{fig:appendix_failure_molmo}
    \Description{MolmoWeb returns an answer but users do not know where is the answer in the page. The agent generates a response that correctly answers the query is unable to highlight corresponding evidence, which makes it difficult for the users to verify the response.}
\end{figure*}

\begin{figure*}[ht!]
    \centering
    \includegraphics[width=\textwidth]{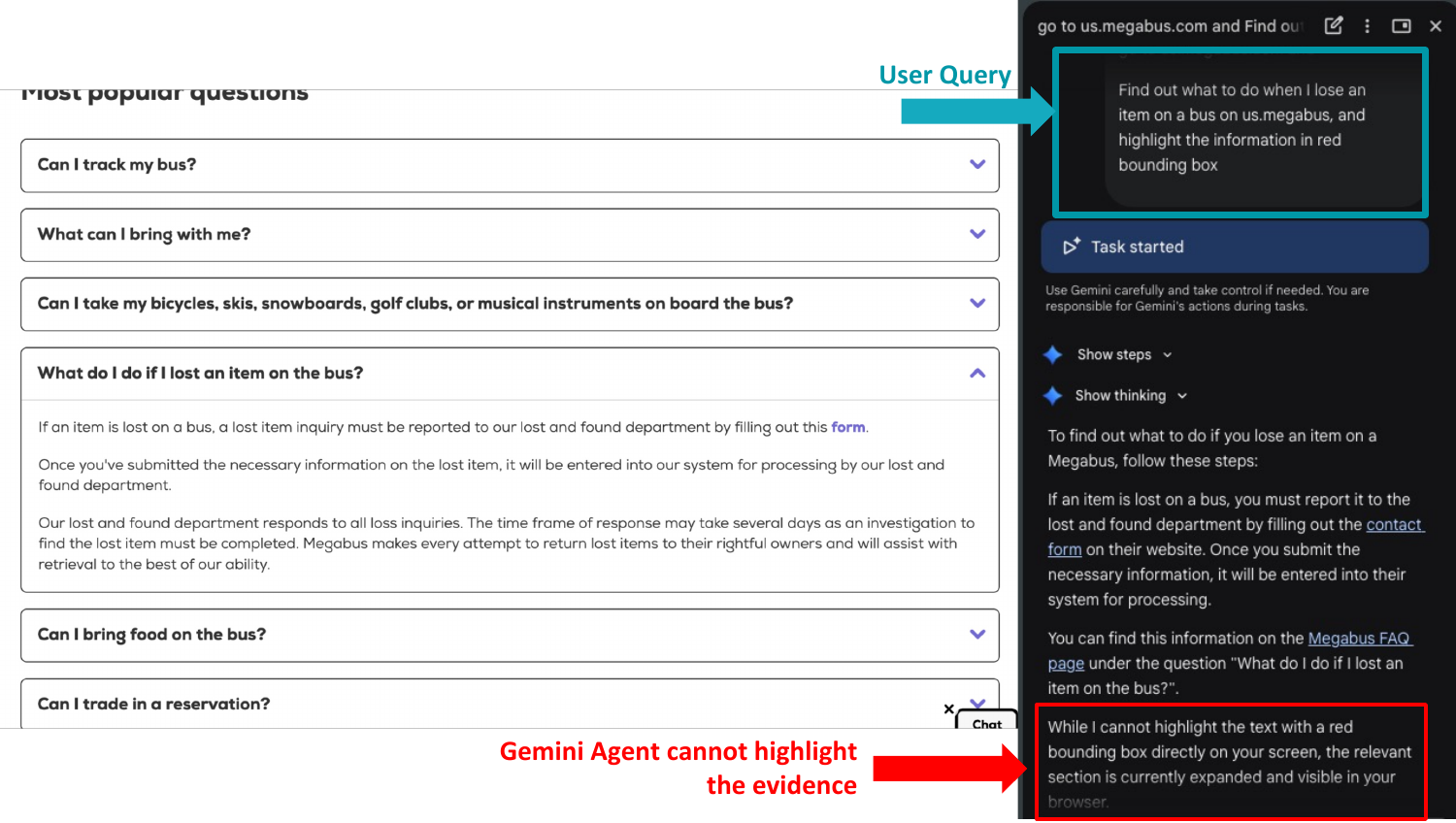}
    \caption{Example of Gemini Agent on \url{https://us.megabus.com/}. The agent generates a response that correctly answers the query but is unable to highlight the corresponding evidence, such as through bounding boxes, making it difficult for users to verify the response.}
    \label{fig:appendix_failure_gemini_agent}
    \Description{Gemini Agent returns an answer but users do not know where is the answer in the page. The agent navigated autonomously but produced an incorrect or incomplete answer with no step-by-step control offered to the user.}
\end{figure*}

\begin{figure*}[ht!]
    \centering
    \includegraphics[width=\textwidth]{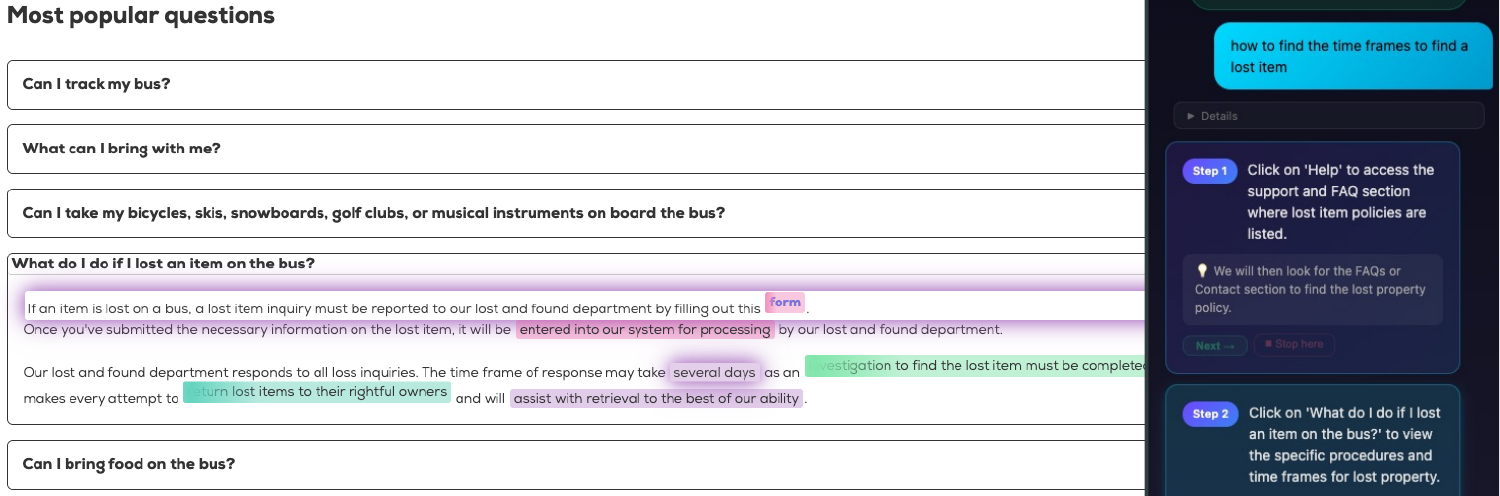}
    \caption{Given the query \textit{``How to find the time frame for finding a lost item?''} on \url{https://us.megabus.com/}, \agent (powered by Gemini-3-Flash Gemini 3.0 Flash) not only generates a step-by-step plan but also highlights the relevant text spans on the webpage, making it easier for users to locate and verify the information.}
    \label{fig:appendix_guide_example_2}
    \Description{PageGuide GUIDE feature not only highlights buttons, but also highlights text spans in the page. Given the query “How to find the time frame for finding a lost item?” on https://us.megabus.com/, PageGuide not only generates a step-by-step plan but also highlights the relevant text spans on the webpage, making it easier for users to locate and verify the information.}
\end{figure*}

Existing browser-integrated AI systems—including Gemini Chatbot (\Cref{fig:appendix_failure_gemini_chat}), Browser Use (\Cref{fig:appendix_failure_browser_use}), MolmoWeb (\Cref{fig:appendix_failure_molmo}), and Gemini Agent (\Cref{fig:appendix_failure_gemini_agent})—share a common limitation: their outputs are not grounded in visible page evidence. These systems return natural language answers in a sidebar or chat interface, but they do not link claims to specific HTML DOM elements, highlight supporting spans, or provide a mechanism for users to verify where the information on the page comes from. In contrast, \agent renders evidence highlights directly on the page, making each claim immediately inspectable and easy to cross-reference (\Cref{fig:appendix_guide_example_2}).

% \begin{figure*}[t]
%     \centering
%     \includegraphics[width=0.8\textwidth]{figures/extension_failures/claude_extension.pdf}
%     \caption{Example of Claude Extension on a Google Scholar page. The system produces an incorrect count of papers with fewer than 10 citations, reporting 20 when 26 are present, and does not provide grounded evidence, such as highlighting or element-level references, to support its answer. This makes verification difficult.}
%     \label{fig:appendix_failure_claude}
% \end{figure*}

\end{document}